\documentclass[preprint, superscriptaddress,showpacs,nofootinbib]{revtex4-1}
\usepackage{amsmath}
\usepackage{graphicx}
\usepackage{amssymb}
\usepackage[colorlinks,citecolor=blue]{hyperref}

\usepackage{color}

\newcommand{\tev}{\, {\rm TeV}}
\newcommand{\gev}{\, {\rm GeV}}
\newcommand{\beq}{\begin{equation}}
\newcommand{\eeq}{\end{equation}}
\newcommand{\bea}{\begin{eqnarray}}
\newcommand{\eea}{\end{eqnarray}}
\newcommand{\nn}{\nonumber}

\newcommand{\cL}{{\cal L}}

\allowdisplaybreaks[4]

\begin{document}

\title{The Diphoton Excess, Low Energy Theorem and the 331 Model}

\author{Qing-Hong Cao}
\email{qinghongcao@pku.edu.cn}
\affiliation{Department of Physics and State Key Laboratory of Nuclear Physics and Technology, Peking University, Beijing 100871, China}
\affiliation{Collaborative Innovation Center of Quantum Matter, Beijing 100871, China}
\affiliation{Center for High Energy Physics, Peking University, Beijing 100871, China}

\author{Yandong Liu}
\email{ydliu@pku.edu.cn}
\affiliation{Department of Physics and State Key Laboratory of Nuclear Physics and Technology, Peking University, Beijing 100871, China}

\author{Ke-Pan Xie}
\email{kpxie@pku.edu.cn}
\affiliation{Department of Physics and State Key Laboratory of Nuclear Physics and Technology, Peking University, Beijing 100871, China}

\author{Bin Yan}
\email{binyan@pku.edu.cn}
\affiliation{Department of Physics and State Key Laboratory of Nuclear Physics and Technology, Peking University, Beijing 100871, China}

\author{Dong-Ming Zhang}
\email{zhangdongming@pku.edu.cn}
\affiliation{Department of Physics and State Key Laboratory of Nuclear Physics and Technology, Peking University, Beijing 100871, China}

\begin{abstract}
We interpret the diphoton anomaly as a heavy scalar $H_3$ in the so-called 331 model. The scalar is responsible for breaking the $SU(3)_C\otimes SU(3)_L\otimes U(1)_X$ gauge symmetry down to the standard model electroweak gauge group. It mainly couples to the standard model gluons and photons through quantum loops involving heavy quarks and leptons. Those quarks and leptons, in together with the SM quarks and leptons, form the fundamental representation of the 331 model. We use low energy theorem to calculate effective couplings of $H_3gg$, $H_3\gamma\gamma$, $H_3ZZ$, $H_3WW$ and $H_3Z\gamma$. The analytical results can be applied to new physics models satisfying the low energy theorem. We show that the heavy quark and lepton contribution cannot produce enough diphoton pairs. It is crucial to include the contribution of charged scalars to explain the diphoton excess. The extra neutral $Z^\prime$ boson could also explain the 2 TeV diboson excess observed at the LHC Run-I.  
\end{abstract}

\maketitle

\section{Introduction}
A diphoton resonance around 750~GeV was reported by the ATLAS and CMS collaborations~\cite{ATLAS-CONF-2015-081, CMS:2015dxe} based on 3.2 and 2.6 ${\rm fb}^{-1}$ of data collected at 13 TeV LHC. ATLAS observed a local significance of 3.9$\sigma$ corresponding to a cross section of about $10\pm 3~{\rm fb}$, while CMS found 2.6$\sigma$ with a smaller cross section of $6 \pm 3~{\rm fb}$. And the large excess around 750~GeV is still confirmed by a recent study from ATLAS and CMS~\cite{ATLAS-CONF-2016-018, CMS-PAS-EXO-16-018}. The resonance is likely to be a new scalar in new physics (NP) model beyond the Standard Model (SM) of particle physics. It has drawn a lot of interests in the field~\cite{Backovic:2015fnp,Harigaya:2015ezk,Mambrini:2015wyu,Angelescu:2015uiz,Pilaftsis:2015ycr,Franceschini:2015kwy,
DiChiara:2015vdm,Low:2015qep,Bellazzini:2015nxw,Ellis:2015oso,McDermott:2015sck,Higaki:2015jag,Gupta:2015zzs,Petersson:2015mkr,
Molinaro:2015cwg,Nakai:2015ptz,Buttazzo:2015txu,Bai:2015nbs,Aloni:2015mxa,Falkowski:2015swt,Csaki:2015vek,Agrawal:2015dbf,Ahmed:2015uqt,
Chakrabortty:2015hff,Bian:2015kjt,Curtin:2015jcv,Fichet:2015vvy,Chao:2015ttq,Demidov:2015zqn,No:2015bsn,Becirevic:2015fmu,
Cox:2015ckc,Kobakhidze:2015ldh,Matsuzaki:2015che,Cao:2015pto,Dutta:2015wqh,Benbrik:2015fyz,Megias:2015ory,Carpenter:2015ucu,
Bernon:2015abk,Alves:2015jgx,Gabrielli:2015dhk,Chao:2015nsm,Arun:2015ubr,Han:2015cty,Chang:2015bzc,Chakraborty:2015jvs,
Ding:2015rxx,Han:2015dlp,Han:2015qqj,Luo:2015yio,Chang:2015sdy,Bardhan:2015hcr,Feng:2015wil,Antipin:2015kgh,Wang:2015kuj,Cao:2015twy,Huang:2015evq,
Liao:2015tow,Heckman:2015kqk,Bi:2015uqd,Cho:2015nxy,Cline:2015msi,Bauer:2015boy,Barducci:2015gtd,Boucenna:2015pav,Murphy:2015kag,
Hernandez:2015ywg,Dey:2015bur,Pelaggi:2015knk,deBlas:2015hlv,Belyaev:2015hgo,Dev:2015isx,Gu:2015lxj,Cvetic:2015vit,
Altmannshofer:2015xfo,Cao:2015xjz,Chakraborty:2015gyj,Badziak:2015zez,Patel:2015ulo,Moretti:2015pbj,Huang:2015rkj,Hall:2015xds,
    Casas:2015blx,Zhang:2015uuo,Liu:2015yec,Cheung:2015cug,Das:2015enc,Craig:2015lra,Davoudiasl:2015cuo,Allanach:2015ixl,Salvio:2015jgu}. Also, an excess of diboson events around 2~TeV was reported in LHC Run-I data~\cite{Aad:2015owa,Khachatryan:2014hpa,Khachatryan:2014gha}. That has been explained in terms of extra gauge bosons such as additional $Z^\prime$ or $W^\prime$ bosons in several NP models with extended gauge structures. New physics models that could explain both anomalies need to have two new ingredients, extra heavy scalars and gauge bosons. The simplest case is the new scalar is singlet under the SM gauge group which requires the model has vector-like fermions in order to couple to the diphoton at the loop-level. Many papers add the singlet scalar and vector-like fermions by hand, while in the so-called 331 model with the $SU(3)_C\otimes SU(3)_L\otimes U(1)_X$ gauge group~\cite{Frampton:1992wt, Pisano:1991ee} these additional particles naturally exist.
There are many versions of the 331 model~\cite{Dias:2003zt,Dias:2003iq,Diaz:2004fs,Dias:2004dc,Dias:2004wk,Dias:2005jm,Dias:2005yh,Dias:2011sq,Ochoa:2005ch} which in general share nice features as follows: i) the anomaly cancellation and the QCD asymptotic freedom require three generation fermions; ii) the Peccei-Quinn (PQ) symmetry~\cite{Peccei:1977hh,Peccei:1977ur} which can solve the strong CP problem is a natural result of gauge invariance in the 331 model~\cite{Dias:2003zt,Dias:2003iq}, etc.  Refs.~\cite{Boucenna:2015pav,Hernandez:2015ywg} studied the diphoton excess in the framework of the 331 gauge symmetry in a specific version of the 331 model while in our paper we consider four possible versions of this model. Also, the trinification model of $SU(3)_L\otimes SU(3)_R\otimes SU(3)_C$, which contains the 331 model as subgroup, has been considered recently in Ref.~\cite{Pelaggi:2015knk} to address on the diphoton anomaly. 

The 331 model consists of very rich particle spectra. For instance, there will be five new gauge bosons, three new heavy quarks, three new leptons, and six new scalars. That has very rich collider phenomenology at the LHC. There are many versions of the 331 model which can be characterized by a parameter called $\beta$. Models with different $\beta$ have new particles with different electric charges. The 331 model has been studied in details by Buras {\it et al} in Ref.~\cite{Buras:2012dp}. In this work we follow the notation in Ref.~\cite{Buras:2012dp} and  consider different versions of the 331 model with $\beta=\pm \sqrt{3}$ and $\pm 1/\sqrt{3}$ to explain the 750~GeV diphoton and 2~TeV diboson excesses. 

The paper is organized as follows. In Sec.~II, we briefly review the 331 model and  introduce the main ingredients of the 331 model needed to explain the diphoton and diboson excesses at the LHC. In Sec.~III we explain how the 331 model could explain the diphoton and diboson signal. Finally, we conclude in Sec. IV.

\section{The Model}

In the 331 model considered in this work, the right-handed fermion fields are treated as singlets of the $SU(3)_L$ group. While the left-handed quark fields of the first two generations are chosen to be in the triplet representation of the $SU(3)_L$ group and the left-handed quarks of the third generation are in the the anti-triplet representation of $SU(3)_L$. Three new quarks ($D$, $S$, $T$) and three new leptons ($E_e$, $E_\mu$, $E_\tau$) are added to form the triplet and anti-triplet. The quark fields are
\beq
q_{1L}=\left(\begin{array}{c}
u  \\
d  \\
D  \\
\end{array}\right)_L, \quad
q_{2L}=\left(\begin{array}{c}
c  \\
s  \\
S  \\
\end{array}\right)_L, \quad
q_{3L}=\left(\begin{array}{c}
b  \\
-t \\
T  \\
\end{array}\right)_L.
\eeq
Note that the $t$ and $b$ assignment is different from the SM as a result of requiring $q_{3L}$ being an anti-triplet. All the left-handed lepton fields of the three generations are all treated as anti-triplets to guarantee the gauge anomaly cancellation, which requires equal numbers of triplets and anti-triplets. The lepton fields are given by
\beq
l_{1L}=\left(\begin{array}{c}
e  \\
 -\nu_e   \\
E_e  \\
\end{array}\right)_L, \quad
l_{2L}=\left(\begin{array}{c}
\mu  \\
 -\nu_\mu   \\
E_\mu  \\
\end{array}\right)_L, \quad
l_{3L}=\left(\begin{array}{c}
\tau  \\
 -\nu_\tau   \\
E_\tau  \\
\end{array}\right)_L, 
\eeq
The extra minus signs in front of top quark and neutrinos are to generate the same Feynman vertices as in the SM.

The symmetry breaking pattern of the 331 model is 
\beq
SU(3)_L\times U(1)_X\rightarrow SU(2)_L\times U(1)_Y\rightarrow U(1)_Q, 
\eeq
which is realized by introducing three scalar triplets $\rho$, $\eta$ and $\chi$
\beq
\rho= \left(\begin{array}{c}
\rho^+  \\
 \rho^0   \\
\rho^{-Q_V}  \\
\end{array}\right) \quad
\eta= \left(\begin{array}{c}
\eta^0  \\
 \eta^-   \\
\eta^{-Q_Y}  \\
\end{array}\right) \quad
\chi= \left(\begin{array}{c}
\chi^{Q_Y}  \\
 \chi^{Q_V}   \\
\chi^0  \\
\end{array}\right), \label{1}
\eeq
where the charges of $Q_V$ and $Q_Y$ are to be determined below. 
The vacuum expectation values (vevs) of $\rho$, $\eta$ and $\chi$ are chosen as 
\beq
\left<\rho\right>=\frac{1}{\sqrt{2}} \left(\begin{array}{c}
0  \\
v_1   \\
0  \\
\end{array}\right)\quad
\left<\eta\right>=\frac{1}{\sqrt{2}} \left(\begin{array}{c}
v_2  \\
0   \\
0  \\
\end{array}\right)\quad
\left<\chi\right>=\frac{1}{\sqrt{2}} \left(\begin{array}{c}
0  \\
0   \\
v_3  \\
\end{array}\right) .
\eeq
It is necessary to mention that the scalar in higher dimensional representation is also possible. However it is impossible to write the Yukawa interaction of the sextet scalar and quarks in a gauge invariant way~\cite{Buras:2012dp} which is essential for the production of the neutral scalar at LHC. Therefore we only consider the three triplets.
At the first step of the symmetry breaking, $\chi$ is introduced to break $SU(3)_L\times U(1)_X$ to $SU(2)_L\times U(1)_Y$ at a very large scale $v_3$, typically at $\tev$ scale. At the second step, we use $\rho$ and $\eta$ to break $SU(2)_L\times U(1)_Y$ down to the residual $U(1)_Q$ electromagnetic symmetry at the weak scale, i.e. $v_1 \sim v_2 \sim m_W$. Thus, we have $v_3$ $\gg$ $v_{1,2}$.
After the first step of symmetry breaking, the hypercharge generator is a linear combination of $X$ and $T^8$ as $Y=\beta T^8+X$, where $T^8$ is diagonal generator of $SU(3)_L$ and $X$ is the generator of $U(1)_X$. The electric charge generator is then given by 
\beq
Q=T^3+\beta T^8+X.
\eeq
Table~\ref{tbl:charge} shows electric charges of new particles. There are exotic quarks (i.e. $D, S, T$) 
for $\beta=\pm \sqrt{3}$ whose electric charges are 4/3 or 5/3~\cite{Cabarcas:2008ys}. For $\beta=1/\sqrt{3}$, the heavy leptons $E$ are neutral and can be identified with right-handed neutrino. By introducing additional sextet scalar~\cite{Ky:2005yq} one can generate small neutrino mass through the seesaw mechanism.

\begin{table}
\caption{\it Electric charges of new particles for different choices of $\beta$.}
\label{tbl:charge}
\begin{tabular}{c | c |c |c| c| c }\hline
particles & $Q(\beta)$ & $\beta=-\frac{1}{\sqrt{3}}$ & $\beta=\frac{1}{\sqrt{3}}$ & $\beta=-\sqrt{3}$ & $\beta=\sqrt{3}$ \\ \hline \hline
$D,S$ & $\frac{1}{6}-\frac{\sqrt{3}\beta}{2}$ & $\frac{2}{3}$ & $-\frac{1}{3}$ & $\frac{5}{3}$ & $-\frac{4}{3}$ \\ \hline
$T$ & $\frac{1}{6}+\frac{\sqrt{3}\beta}{2}$ & $-\frac{1}{3}$ & $\frac{2}{3}$ & $-\frac{4}{3}$ & $\frac{5}{3}$ \\ \hline
$E$ & $-\frac{1}{2}+\frac{\sqrt{3}\beta}{2}$ & $-1$ & $0$ & $-2$ & $1$ \\ \hline
$V$ & $-\frac{1}{2}+\frac{\sqrt{3}\beta}{2}$ & $-1$ & $0$ & $-2$ & $1$ \\ \hline
$Y$ & $\frac{1}{2}+\frac{\sqrt{3}\beta}{2}$ & $0$ & $1$ & $-1$ & $2$ \\ \hline
$H_V$ & $-\frac{1}{2}+\frac{\sqrt{3}\beta}{2}$ & $-1$ & $0$ & $-2$ & $1$ \\ \hline
$H_Y$ & $\frac{1}{2}+\frac{\sqrt{3}\beta}{2}$ & $0$ & $1$ & $-1$ & $2$ \\ \hline
$H_W$ & $1$ & $1$ & $1$ & $1$ & $1$ \\ \hline
 \end{tabular}

\end{table}

The scalar potential of the three scalar triplets is
\bea
V_{H}&=&\mu_1^2 (\rho^\dagger \rho)+
 \mu_2^2 (\eta^\dagger \eta)+\mu_3^2 (\chi^\dagger \chi)+ \lambda_1 (\rho^\dagger \rho)^2+
 \lambda_2 (\eta^\dagger \eta)^2+\lambda_3 (\chi^\dagger \chi)^2  \nn \\
 &+& \lambda_{12}(\rho^\dagger \rho)(\eta^\dagger \eta)+
\lambda_{13}(\rho^\dagger \rho)(\chi^\dagger \chi)+
\lambda_{23}(\eta^\dagger \eta)(\chi^\dagger \chi)  \nn \\
 &+& \lambda^\prime_{12}(\rho^\dagger \eta)(\eta^\dagger \rho)+
\lambda^\prime_{13}(\rho^\dagger \chi)(\chi^\dagger \rho)+
\lambda^\prime_{23}(\eta^\dagger \chi)(\chi^\dagger \eta ) + \sqrt{2} f \left(\epsilon_{ijk}\rho^i \eta^j \chi^k+{\rm h.c.}\right)\label{VHiggs},
\eea
where $\lambda^{(\prime)}_{i(j)}$ denotes the dimensionless parameter while $\mu_i$ and the parameter $f$ 
has a mass dimension. For simplicity, we assume $f$ is proportional to $v_3$ such that the theory has no other scale, i.e. $f \equiv k v_3$ with $k\sim O(1)$. Three CP-even scalars and one CP-odd scalar ($A$) remain after symmetry breaking. 
The mass mixing matrix of the CP-even scalars is
\beq
M_{H}^2=
\left(
\begin{array}{ccc}
 2 \lambda _1 v_1^2+\frac{f v_2 v_3}{v_1} & -f v_3+v_1 v_2 \lambda _{12} & -f v_2+v_1 v_3 \lambda _{13} \\
 -f v_3+v_1 v_2 \lambda _{12} & 2 \lambda _2 v_2^2+\frac{f v_1 v_3}{v_2} & -f v_1+v_2 v_3 \lambda _{23} \\
 -f v_2+v_1 v_3 \lambda _{13} & -f v_1+v_2 v_3 \lambda _{23} & 2 \lambda _3 v_3^2+\frac{f v_1 v_2}{v_3} \\
\end{array}
\right)\label{Neutral even mass mixing}.
\eeq
In the limit of $v_3\gg v_{1,2}$, the three eigenvalues are
\beq
M_h^2=\lambda_1 v_1^2+\lambda_2 v_2^2,\quad
M_{H_2}^2=\frac{\left(v_1^2+v_2^2\right)}{v_1 v_2}fv_3,\quad
M_{H_3}^2=2 \lambda _3 v_3^2.
\eeq
Here $h$, $H_2$ and $H_3$ are three mass eigenstates of the neutral scalars, which are related with the weak eigenstates by a rotation matrix
\beq
\left(\begin{array}{c}
h\\
H_2\\
H_3\\
\end{array}\right)=
\left(\begin{array}{c c c}
\frac{v_1}{\sqrt{v_1^2+v_2^2}} & -\frac{v_2}{\sqrt{v_1^2+v_2^2}} & 0\\
\frac{v_2}{\sqrt{v_1^2+v_2^2}} & \frac{v_1}{\sqrt{v_1^2+v_2^2}} & 0\\
0 & 0 & 1\\
\end{array}\right)
\left(\begin{array}{c}
\xi_\rho\\
\xi_\eta\\
\xi_\chi\\
\end{array}\right),
\eeq
where $\xi_\rho$, $\xi_\eta$ and $\xi_\chi$ are weak eigenstates of the CP-even scalars.
One of the three CP-even scalars is identified as the 125~GeV Higgs boson ($h$) while the other two ($H_2$ and $H_3$) are much heavier. As a result of the two-step symmetry breaking, the $H_3$ scalar is mainly from the $\chi$ triplet while the $h$ and $H_2$ scalars  are from the $\rho$ and $\eta$ triplets. The mixing between the $H_3$ scalar and the Higgs boson $h$ is suppressed such that decays of the $H_3$ scalar into the SM particles are negligible. On the other hand, the $H_2$ scalar has a sizable mixing with the Higgs boson and then could decay into a pair of SM fermions, $W$ and $Z$, and Higgs bosons, if kinematically allowed. 

The CP-odd scalar $A$ mainly arises from $\rho$ and $\eta$ as the imaginary part of the neutral component in the $\chi$ triplet is eaten by the new $Z^\prime$ boson. The scalar $A$ could decay into the SM quarks and leptons directly. There are three charged scalars after symmetry breaking, named as $H_W$, $H_V$ and $H_Y$, whose electric charges are $\pm1$, $\pm Q_V$ and $\pm Q_Y$, respectively. See Table~\ref{tbl:charge} for the scalar charges for different values of $\beta$.  

In addition to the $W$-boson and the $Z$-boson, four charged gauge bosons and one neutral gauge boson appear after symmetry breaking. The four charged gauge bosons carry the same charge as the charged scalars; see Table~\ref{tbl:charge}. We thus name them as $Y^{\pm Q_Y}$ and $V^{\pm Q_V}$ for simplicity. All the new gauge bosons achieve their masses after the first step of symmetry breaking and their masses are naturally around $v_3$. Below we summarize the mass spectra of the vector bosons and scalars in the limit of $v_3\gg v_{1,2}$: 
\begin{itemize}
\item Gauge bosons
\begin{align}
&M_{W}=\frac{1}{2}g\sqrt{v_1^2+v_2^2},\quad &&M_{Y}=M_{V}=\frac{1}{2}gv_3, \nn\\ 
&M_{Z}=\frac{1}{c_W} M_{W},\quad &&M_{Z^\prime} = \frac{2c_W}{\sqrt{3\left[1-(1+\beta^2)s_W^2\right]}}M_{Y},
\end{align}
where $c_W\equiv \cos\theta_W$ and $s_W\equiv \sin\theta_W$ with $\theta_W$ being the Weinberg mixing angle.
  \item Scalars
\begin{align}
&M_{H_W}\approx M_{H_2}\approx M_{A}\approx v_3\sqrt{\frac{k(v_1^2+v_2^2)}{v_1 v_2}}, \quad  &&M_{H_3} = \sqrt{2 \lambda_3} v_3, \nn\\
&M_{H_Y} = v_3 \sqrt{k \frac{v_1}{v_2} + \frac{\lambda_{23}^\prime}{2}}, \quad &&M_{H_V}=v_3 \sqrt{k \frac{v_2}{v_1} + \frac{\lambda_{13}^\prime}{2}},
\end{align}
where $k\equiv f/v_3 \sim \mathcal{O}(1)$ is understood. The degeneracy of $A$, $H_{2}$ and $H_W$ is owing to the remaining $SU(2)$ symmetry after the first step of  symmetry breaking. 
\end{itemize}
The masses of new fermions ($\mathcal{F}$) originate from the following Yukawa Lagrangian,
\bea
-\cL^{\mathcal {F}}_{Yuk}&=&y^J_{ik}\bar q_{iL}\chi J_{kR}+y^J_{33}\bar q_{3L}\chi^* J_{3R} +y^E_{mn}\bar l_{mL}\chi^* E_{nR}+h.c.~~,
\eea
where $i$ ($k$) run from 1 to 2 while $j$ ($m$, $n$) run from 1 to 3. $J_{kR}$ refers to the right-handed heavy quarks $D$ and $S$, $J_{3R}$ to $T$, and $E_{nR}$ represents the right-handed heavy leptons $E_{e,\mu,\tau}$. 
Note that $\chi$ can only give mass to the third component of the triplet fermions (i.e. $E_e, E_\mu, E_\tau, D, S, T$), therefore, there is no direct couplings of $\chi$ to the SM particles.
After the first step of the symmetry breaking, the new fermions obtain their masses from the vacuum expectation value of $\chi$. For simplicity, we assume the mass matrix of the new fermions is diagonal, leading to the following masses,
\begin{align}
M_D&=\frac{y^J_{11}}{\sqrt{2}}v_3 &M_S=\frac{y^J_{22}}{\sqrt{2}}v_3 &&M_T=\frac{y^J_{33}}{\sqrt{2}}v_3 \\
M_{E_e}&=\frac{y^E_{11}}{\sqrt{2}}v_3 &M_{E_\mu}=\frac{y^E_{22}}{\sqrt{2}}v_3 &&M_{E_\tau}=\frac{y^E_{33}}{\sqrt{2}}v_3 
\end{align}
For illustration we choose benchmark parameters
\beq
M_{E_e}=M_{E_\mu}=M_{E_\tau}=M_D=M_S=M_T=800\gev.
\eeq

\section{Diphoton excess}

\begin{figure}
\includegraphics[scale=0.4]{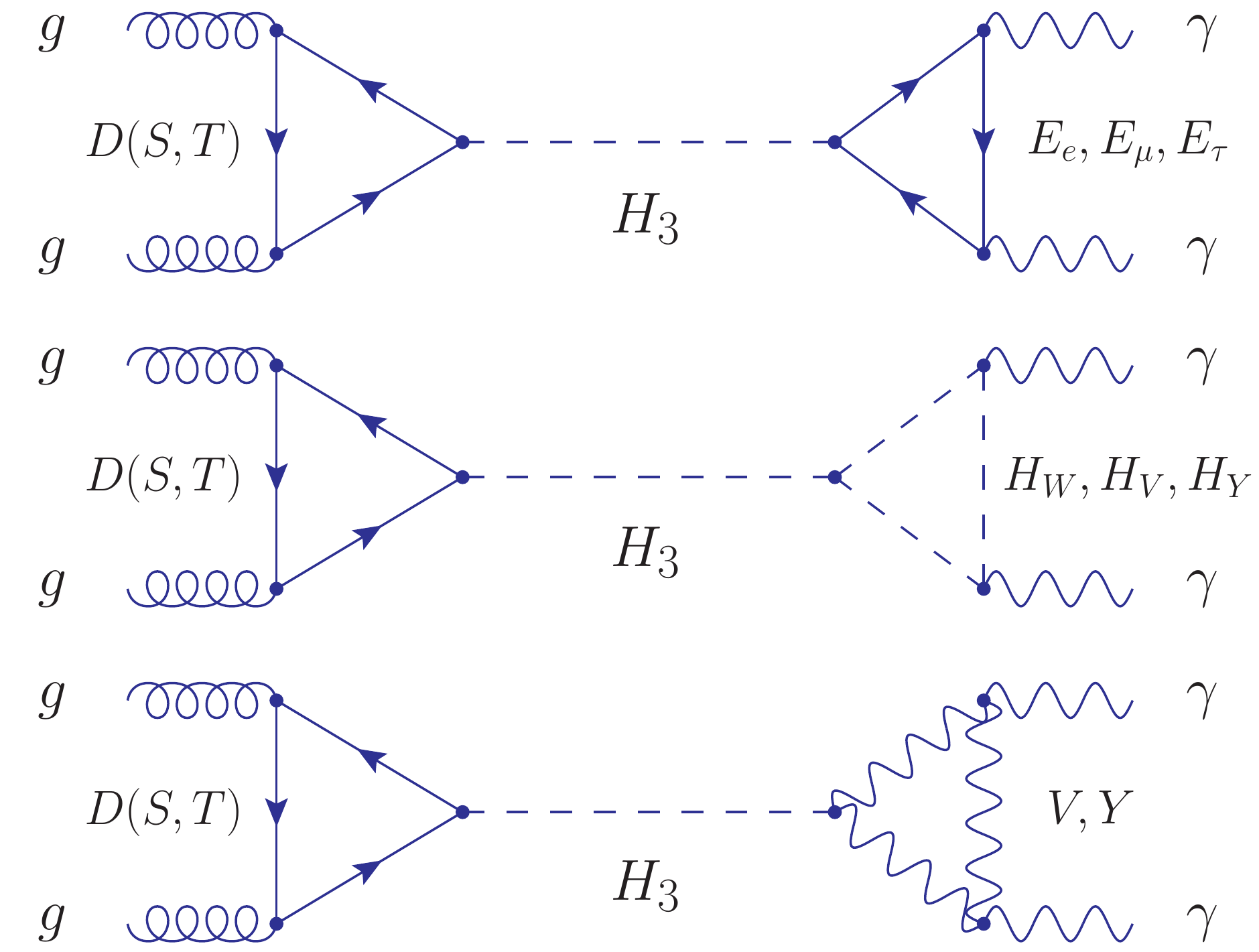}
\caption{The Feynman diagrams for $gg\to H\to \gamma\gamma$.  }
\label{diphoton}
\end{figure}

To explain the diphoton excess, the cross section $\sigma(pp\to X \to \gamma\gamma)$ is estimated roughly as $(8\pm 5)~{\rm fb}$~\cite{ATLAS-CONF-2015-081, CMS:2015dxe} where $X$ denotes the possible resonance state. In the 331 model the $H_3$ scalar arises  mainly from the $\chi$ triplet which decouples from the electroweak sector in the SM. The $H_3$ scalar does not couple to the SM particles directly. However, it could be produced through the heavy quark loops and also decays via the loops of new heavy charged particles; see Fig.~\ref{diphoton}.  Using the narrow width approximation (NWA), the cross section of $\sigma(pp\to H_3 \to \gamma\gamma)$  at the LHC with energy $\sqrt{s}=13~{\rm TeV}$ could be parameterized as~\cite{Franceschini:2015kwy}
\beq
\sigma(pp\to H_3 \to \gamma\gamma) = \frac{\mathcal{C}_{gg}}{M_{H_3}\Gamma s} \Gamma(H_3\to  gg) \Gamma(H_3\to \gamma\gamma),
\label{eq:master}
\eeq 
where $\mathcal{C}_{gg}=2137$ denotes the dimensionless partonic integral of parton distribution functions, and $M_{H_3}$ is the mass of the $H_3$ scalar resonance. The partial decay widths of $H_3$ are given by
\begin{align}
\Gamma(H_3\to gg) &= \frac{1}{8\pi}c_{gg}^2 M_{H_3},\nn\\
\Gamma\left(H_3\to\gamma\gamma\right)&=\dfrac{1}{64\pi}\left(-c_{\gamma\gamma}^{\mathcal F}+c_{\gamma\gamma}^{\mathcal V}+c_{\gamma\gamma}^{\mathcal S}\right)^2M_{H_3},\nn\\
\Gamma\left(H_3\to ZZ\right)&=\mathcal{P}\left(\dfrac{m_Z^2}{M_{H_3}^2}\right)\dfrac{1}{64\pi}\left(c_{ZZ}^{\mathcal  V}-c_{ZZ}^{\mathcal  F}+c_{ZZ}^{\mathcal  S}\right)^2M_{H_3},\nn\\
\Gamma\left(H_3\to WW\right)&=\mathcal{P}\left(\dfrac{m_W^2}{M_{H_3}^2}\right)\dfrac{1}{32\pi}\left(c_{WW}^{\mathcal  V}-c_{WW}^{\mathcal  F}+c_{WW}^{\mathcal  S}\right)^2M_{H_3},\nn\\
\Gamma\left(H_3\to Z\gamma\right)&=\dfrac{1}{32\pi}\left(1-\dfrac{m_Z^2}{M^2_{H_3}}\right)^3\left(c_{Z\gamma}^{\mathcal  V}-c_{Z\gamma}^{\mathcal F}+c_{Z\gamma}^{\mathcal S}\right)^2M_{H_3},
\label{eq:parwidth}
\end{align}
where $\mathcal{P}(x)=\sqrt{1-4x} \left(1-4x+6x^2\right)$ is a factor correcting for the massive final states in the decay width. The corresponding coefficients denoted by $c_{ij}^k$ can be derived from the Higgs low-energy theorem which will be shown in the following subsections. We have checked the expressions of the coefficients with those from Ref.~\cite{Djouadi:2005gi, Shifman:1979eb, Cao:2009ah}. Note that the total decay width $\Gamma$ can be obtained by summing over the above partial decay widths.

\subsection{Loop-induced $H_3 \to gg$}
The interaction of the $H_3$ scalar to the gluon pair through heavy quarks can be derived from the low energy theorem~\cite{Low:2009di}. In the SM gluon fusion production is induced by the top quark loop~\cite{Djouadi:2005gi},
\beq
c_{gg}^{\rm SM} = \frac{\sqrt{2}\alpha_s}{3\pi} \frac{m_h}{v}
\eeq
It is well-known that this coefficient is related to the top contribution to the gluon two-point function from the Higgs low-energy theorem.
Next we consider the case when the loop contains a pair of vectorlike quarks $(Q^c, Q)$ charged under the electroweak gauge group with the interaction 
\beq
M_Q Q^c Q+y_Q H_3 Q^c Q,
\eeq
then its contribution to the gluon two-point function is~\cite{Cao:2009ah}
\beq
-\frac{1}{4} \left[ 1- \frac{g_s^2}{16\pi^2} b_F^{(3)} \log\frac{M_Q^2(\mathcal{H})}{\mu^2}\right] G^a_{\mu\nu}G^{a~\mu\nu},\label{eq:hgg}
\eeq
where $b_F^{(3)} = 2/3$ is the contribution to the one-loop beta function of QCD from a Dirac fermion and
$M_Q(\mathcal{H}) = M_Q + y_Q \mathcal{H}$ is the mass of the new heavy fermion $Q$ when turning on the scalar as a background field $H_3\to \mathcal{H}$. The $y_Q$ denotes the Yukawa coupling of the $H_3$ scalar to heavy quark $Q$. To obtain the scalar-gluon-gluon coupling, the Higgs low-energy theorem instructs us to expand Eq.~\ref{eq:hgg} to the first order in $\mathcal{H}$~\cite{Low:2009di}:
\beq
c_{gg} = \sum_{Q=D,S,T}\frac{\alpha_s}{3\pi} \frac{M_{H_3}}{M_Q} y_Q.
\eeq
Strictly speaking, the low-energy theorem applies only when the mass of the particle in the loop is much larger than the scalar mass, $M^2_{H_3}/(4M^2_Q) \ll 1$, so that the loop diagram can be approximated by a dimension-five operator.

\subsection{Loop induced $H_3\to \gamma\gamma$}
The decay of $H_3\to \gamma\gamma$ can occur through a set of heavy fermions ($\mathcal{F}$), scalars ($\mathcal{S}$) and vector bosons ($\mathcal{V}$). 
First, the contribution of heavy quarks and leptons to the two-point function  of the photon is 
\bea
c_{\gamma\gamma}^{\mathcal{F}} = \sum_{\mathcal{F}} \frac{2\alpha N_c}{3\pi}Q_{\mathcal F}^2 \frac{M_{H_3}}{M_\mathcal{F}} y_\mathcal{F}
\eea
where $\mathcal{F}$ denotes the heavy quarks ($D,S,T$) and leptons ($E_e,E_\mu,E_\tau$). Here, $Q_{\mathcal F}$ is the electric charge of those heavy fermions and $N_c=3~(1)$ for quarks (leptons) is understood.
We note that there is a strong cancellation between new charged vector-bosons and new fermions, which leads to a small partial width of $H\to\gamma\gamma$. 
 
In the 331 model the $H_3\to\gamma\gamma$ can also be induced by a new set of charged vector-bosons $V$ and $Y$. As the electric charges of $V$ and $Y$ depend on the choice of $\beta$, we introduce a general form of  gauge boson couplings $\kappa g_{WW\gamma}^{\rm SM}$ where $g_{WW\gamma}^{\rm SM}$ is the $WW\gamma$ coupling in the SM. The contribution of  charged vector bosons is simply
\beq
-\frac{1}{4} \left[1-\frac{e^2 \kappa^2}{16\pi^2 }7 \log\frac{M_{\mathcal V}^2(\mathcal{H})}{\mu^2}\right] F_{\mu\nu}F^{\mu\nu},
\eeq
where $\mathcal{V}=V, Y$ and $F_{\mu\nu}$ is the field strength tensor of photon.
Thus, we obtain the contribution of heavy new gauge bosons to the two-point function of photon is
\beq
c_{\gamma\gamma}^{\mathcal V}=\sum_{\mathcal V}\dfrac{7\alpha\kappa^2}{2\pi}\dfrac{M_{H_3}}{M_{\mathcal V}}y_{\mathcal V}.
\eeq
In the SM the $W$ contribution to the decay of $H\to \gamma\gamma$ dominates over the one from the top-quark loop owing to a large beta function coefficient ``7" multiplying the $W$ loop result. 

Another contribution is from the charged scalar $\mathcal{S}$ ($H_W$, $H_V$, $H_Y$) loops. For a general coupling of the photon to a pair  of charged Higgs bosons,
\beq
g_{\gamma \mathcal{S} \mathcal{S}} = \lambda \left(p_1-p_2\right)^\mu,
\eeq
the loop contribution reads as 
\beq
-\frac{1}{4} \left[1+\frac{\lambda^2}{48\pi^2 } \log\frac{M_{\mathcal S}^2(\mathcal{H})}{\mu^2}\right] F_{\mu\nu}F^{\mu\nu}.
\eeq 
In the limit of $v_3\gg v_{1,2}$, the charged scalar masses are proportional to $v_3$, which leads to
\beq
c_{\gamma\gamma}^{\mathcal S}=-\sum_{\mathcal S}\dfrac{\lambda^2}{24\pi^2}\dfrac{M_{H_3}}{M_{\mathcal S}}y_{\mathcal S}.
\eeq
In the 331 model, the mass of charged scalar ($H_W$, $H_V$, $H_Y$) also depends on other parameters of the Higgs potential. The corresponding coefficient is then given by 
\beq
c_{\gamma\gamma}^{\mathcal S}(331)=-\sum_{\mathcal S}\dfrac{\lambda^2}{48\pi^2}\dfrac{\lambda_{\rm eff}v_3}{M_{\mathcal S}^2}M_{H_3},
\eeq
where $\lambda_{\rm eff}$ denotes the effective coupling of $H_3$ to charged scalars and $M_{\mathcal S}$ denotes the mass of charged scalar.

\subsection{Loop induced $H_3\to ZZ/WW$}

For simplicity, we consider general interactions as follows:
\begin{align}
&g_{Z{\mathcal V}{\mathcal V}} = \kappa^\prime g_{ZWW}^{\rm SM}, \quad
&& g_{Z\mathcal{F}\mathcal{F}} = g \gamma^\mu \left(g_L P_L + g_R P_R\right), \nn\\
& g_{Z\mathcal{S}\mathcal{S}} = \lambda^\prime \left(p_1-p_2\right)^\mu, &&g_{Z\mathcal{S}\mathcal{V}} = \eta M_{\mathcal V} g^{\mu\nu}.
\label{eq:zcoupling}
\end{align}
The contribution of the vector boson loop to the two-point function of the $Z$-boson is
\beq
-\frac{1}{4} \left[1-\frac{e^2 \kappa^{\prime\,2}}{16\pi^2 s_W^2 c_W^2}7 c_W^4 \log\frac{M_{\mathcal V}^2(\mathcal{H})}{\mu^2}\right] Z_{\mu\nu}Z^{\mu\nu}.
\label{eq:vzz}
\eeq
We expand Eq.~\ref{eq:vzz} to the first order in $H_3$ to obtain the effective coefficent
\beq
c_{ZZ}^{\mathcal V}=\sum_{\mathcal V}\dfrac{7\alpha\kappa^{\prime\,2}}{2\pi}\dfrac{c_W^2}{s_W^2}\dfrac{M_{H_3}}{M_{\mathcal V}}y_{\mathcal V}.
\eeq
The fermion loop contribution is 
\beq
-\frac{1}{4} \left[1-\frac{e^2 N_c}{24\pi^2 s_W^2}\left(g_L^2 + g_R^2\right) \log\frac{M_{\mathcal  F}^2(\mathcal{H})}{\mu^2}\right] Z_{\mu\nu}Z^{\mu\nu},
\eeq
and
\beq
c_{ZZ}^{\mathcal  F}=\sum_{\mathcal  F}\dfrac{\alpha N_c}{3\pi s_W^2}\left(g_L^2+g_R^2\right)\dfrac{M_{H_3}}{M_{\mathcal F}}y_{\mathcal  F}.
\eeq
The scalar loop contribution is 
\beq
-\frac{1}{4} \left[1+\frac{\lambda^{\prime \, 2}}{48\pi^2 } \log\frac{M_{\mathcal S}^2(\mathcal{H})}{\mu^2} + \frac{\eta^2}{192\pi^2}\log\frac{M_{\mathcal  V}^2(\mathcal{H})}{\mu^2}\right] Z_{\mu\nu}Z^{\mu\nu},
\eeq
and
\beq
c_{ZZ}^{\mathcal  S}=-\sum_{\mathcal  S}\left(\dfrac{\lambda^{\prime \, 2}}{24\pi^2}+\dfrac{\eta^2}{96\pi^2}\right)\dfrac{M_{H_3}}{M_{\mathcal S}}y_{\mathcal S}.
\eeq

The decay of $H_3\to W^+W^-$ can be easily obtained from the result of $H_3\to ZZ$ decay. We consider the couplings between $W$-boson and new particles are same as Eq.~\ref{eq:zcoupling}, then 
\beq
c_{WW}^{\mathcal  V}=c_{ZZ}^{\mathcal V},\quad
c_{WW}^{\mathcal  F}=c_{ZZ}^{\mathcal F},\quad
c_{WW}^{\mathcal  S}=c_{ZZ}^{\mathcal  S}.
\eeq

\subsection{Loop induced $H_3\to Z\gamma$}
The decay of $H_3\to Z\gamma$ can also occur through  a set of heavy fermions, scalars and vector bosons. 
The contribution of the vector boson loop is 
\beq
-\frac{1}{2} \left[1-\frac{7 e^2 c_W}{16\pi^2 s_W }\kappa \kappa^\prime \log\frac{M_{\mathcal V}^2(\mathcal{H})}{\mu^2}\right] F_{\mu\nu}Z^{\mu\nu},
\eeq
the fermion loop contribution is
\beq
-\frac{1}{2} \left[1+ \frac{e^2 N_c Q_{\mathcal  F}}{24\pi^2 s_W}\left(g_L + g_R\right) \log\frac{M_{\mathcal  F}^2(\mathcal{H})}{\mu^2}\right] F_{\mu\nu}Z^{\mu\nu},
\eeq
and the scalar loop contribution is 
\beq
-\frac{1}{2} \left[1-\frac{\lambda \lambda^\prime}{48\pi^2} \log\frac{M_{\mathcal S}(\mathcal{H})}{\mu^2}\right]F_{\mu\nu}Z^{\mu\nu}.
\eeq
We thus obtain
\bea
c_{Z\gamma}^{\mathcal V} &=& \sum_{\mathcal V}\dfrac{7\alpha}{2\pi}\dfrac{c_W}{s_W}\kappa\kappa^\prime\dfrac{M_{H_3}}{M_{\mathcal V}}y_{\mathcal V},\nn\\
c_{Z\gamma}^{\mathcal F} &=& -\sum_{\mathcal F}\dfrac{\alpha N_cQ_{\mathcal  F}}{3\pi s_W }\left(g_L+g_R\right)\dfrac{M_{H_3}}{M_{\mathcal F}}y_{\mathcal  F},\nn\\
c_{Z\gamma}^{\mathcal S} &=& \sum_{\mathcal S}\dfrac{\lambda\lambda^\prime}{24\pi^2}\dfrac{M_{H_3}}{M_{\mathcal S}}y_{\mathcal S}.
\eea

\subsection{The 750 GeV Diphoton excess }

\begin{figure}[b]
\includegraphics[scale=0.3]{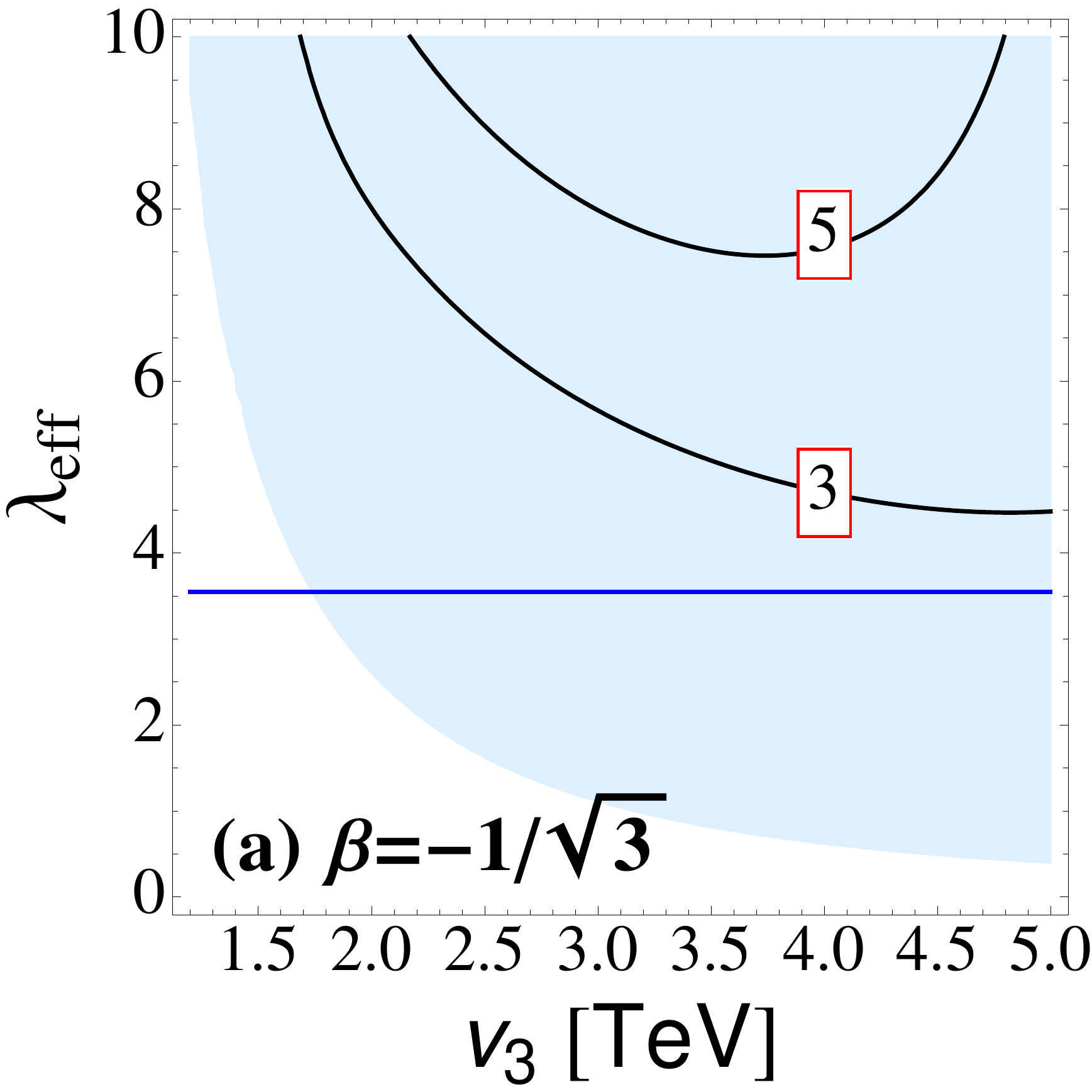}
\includegraphics[scale=0.3]{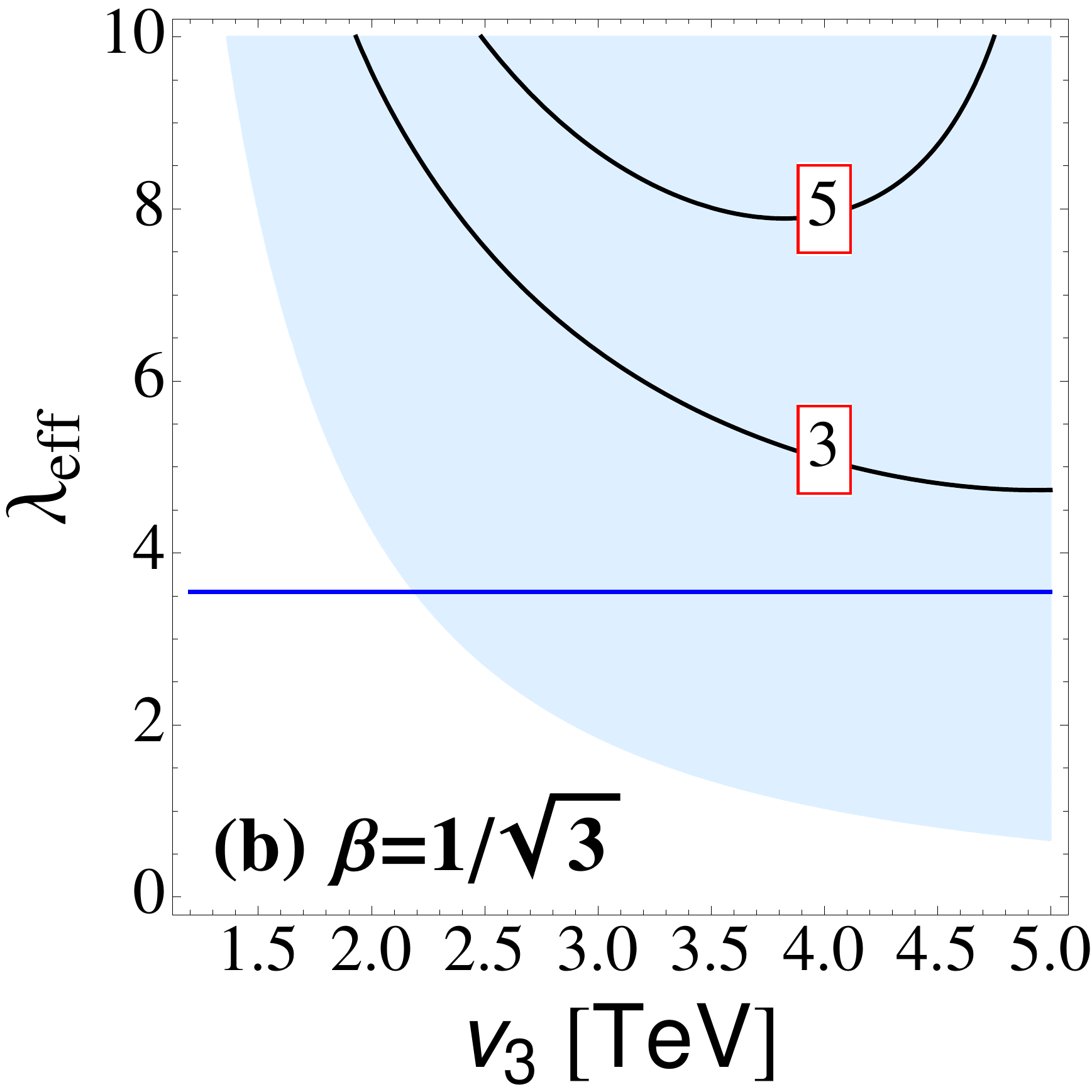}\\
\includegraphics[scale=0.3]{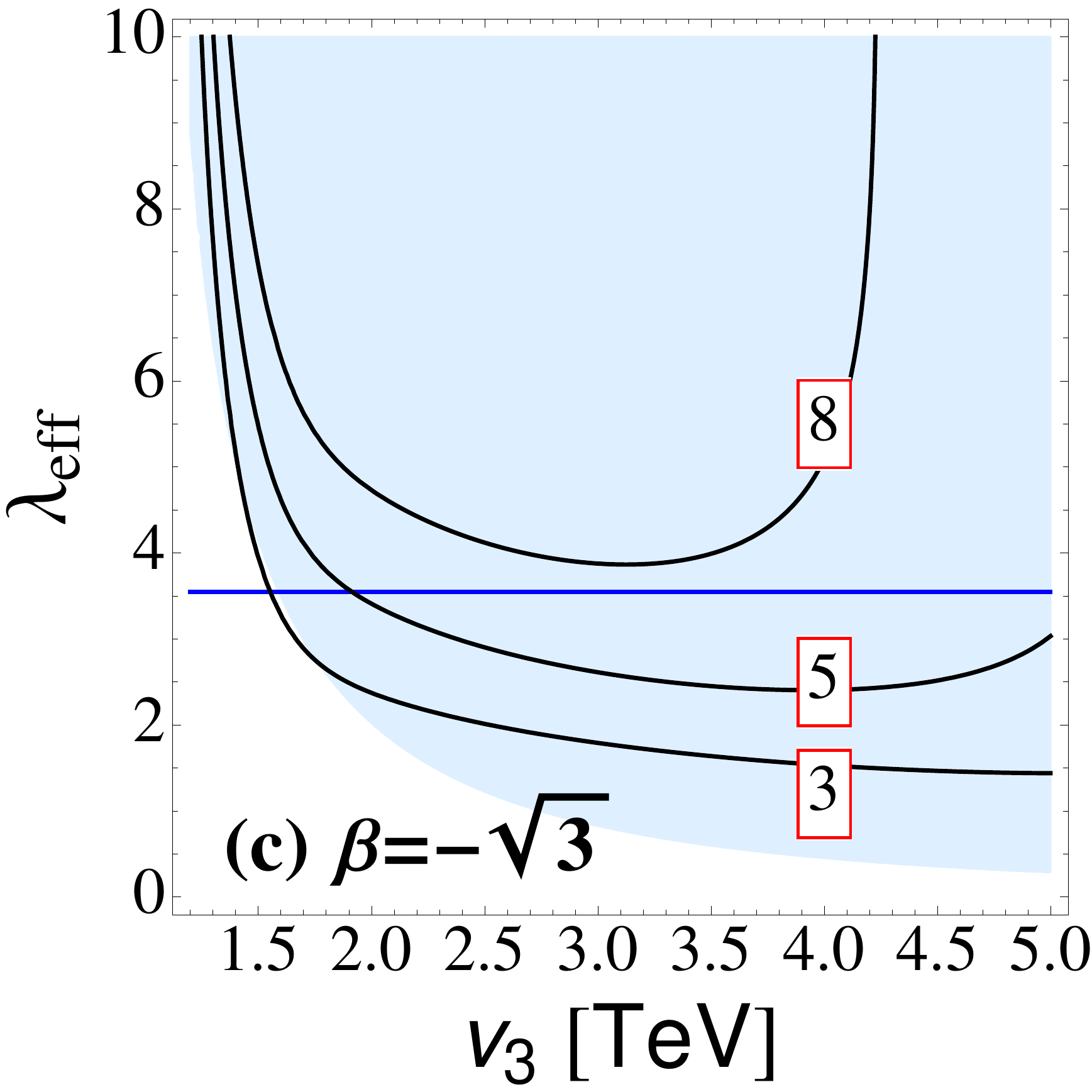}
\includegraphics[scale=0.3]{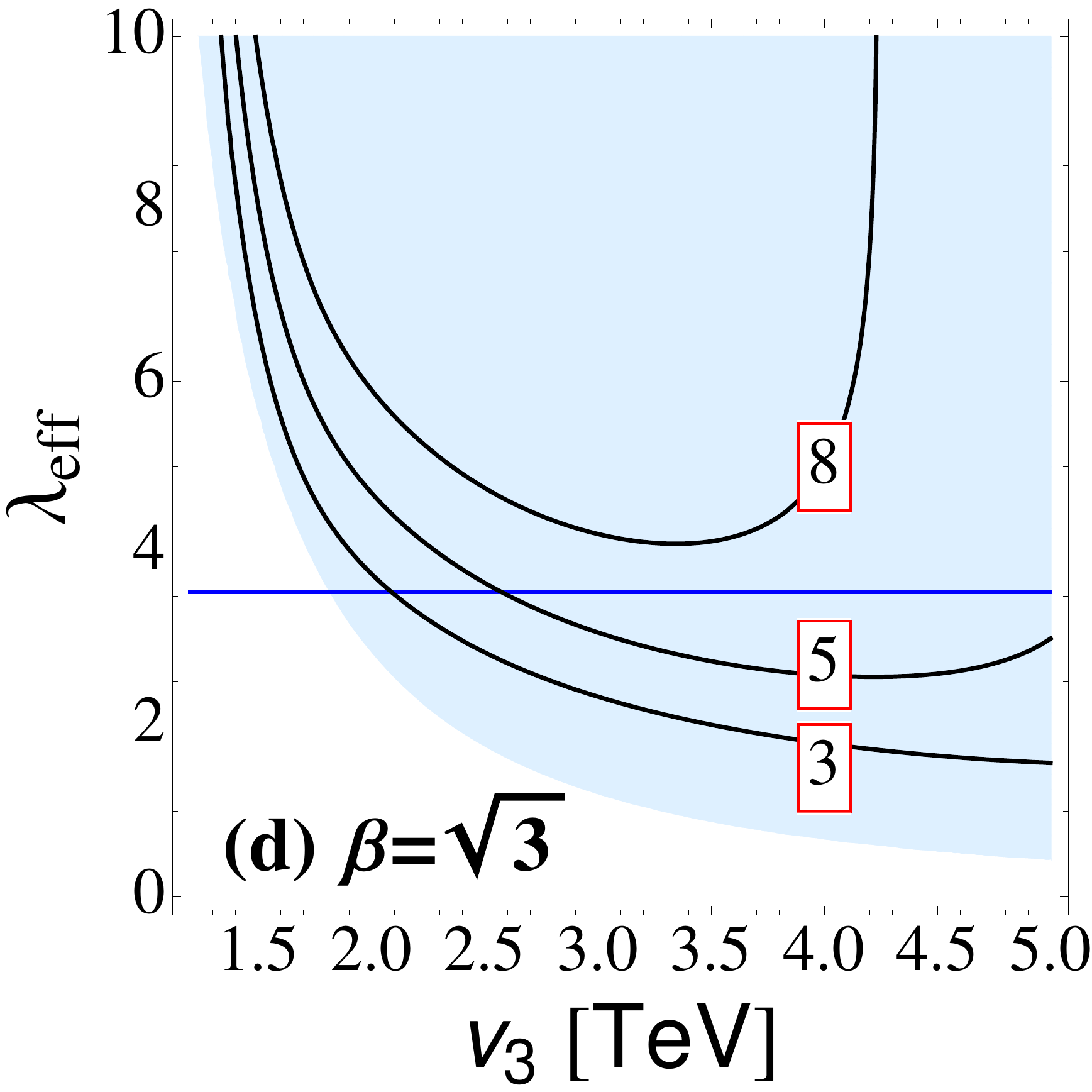}
\caption{\it The contour of $\sigma(pp\to H_3\to \gamma\gamma)$ (fb) in the plane of $\lambda_{eff}$ and $v_3$. The shaded region is consistent with the combined bound of the $WW, ZZ, Z\gamma$, and $gg$ productions at the LHC Run-I while the blue horizontal line represents the perturbation upper limit.  }
\label{fig:diphoton}
\end{figure}

Equipped with all the above analytical results, we calculate all the partial widths of the $H_3$ scalar and the cross section of $pp\to H_3\to \gamma\gamma$ using Eq.~\ref{eq:master}. We choose $M_{H_W}=M_{H_V}=M_{H_Y}=800~{\rm GeV}$ as our benchmark parameters.
Figure~\ref{fig:diphoton} displays the cross section contour of $pp\to H_3\to \gamma\gamma$ at the 13~TeV LHC in the plane of $\lambda_{\rm eff}$ and $v_3$, where $\lambda_{\rm eff}$ in the limit of $v_3\gg v_{1,2}$ is given by
\begin{itemize}
\item $H_3H_WH_W:$ $2k v_2/\sqrt{v_1^2+v_2^2}+\lambda_{13} v_2^2/(v_1^2+v_2^2)+\lambda_{23}$
\item $H_3H_YH_Y:$ $\lambda_{23}+\lambda_{23}^\prime$
\item $H_3H_VH_V:$ $\lambda_{13}+\lambda_{13}^\prime$
\end{itemize}
For simplicity, we use a universal effective coupling $\lambda_{\rm eff}$ to represents the above three couplings. The shaded region represents the parameter space allowed by the combined constraint from the $WW, ZZ, Z\gamma$ and $gg$ productions at the LHC Run-1~\cite{Aad:2014fha}. The di-jet constraint can be satisfied over the entire parameter space shown in Fig.~\ref{fig:diphoton}. The blue horizontal line denotes the perturbative bound $\lambda_{\rm eff}\leq \sqrt{4\pi}$. The cross section for $\beta=\pm \sqrt{3}$ is larger than that for $\beta=\pm 1/\sqrt{3}$ due to the electric charge enhancement of the charged scalars; see Table~\ref{tbl:charge}. For the models of $\beta=\pm 1/\sqrt{3}$, one cannot obtain enough rate of the diphoton rate to explain the excess even with a large $\lambda_{\rm eff}$ close to the perturbative limit. For a small $v_3$, say $v_3\leq 1150~{\rm GeV}$, $M_{V} = M_{Y}\leq M_{H_3}/2$. In such a case the decay channels of $H_3\to VV/YY$ open at the tree level.  That increases the total width of the $H_3$ scalar and reduces the branching ratio of $H_3\to\gamma\gamma$ significantly. A fairly large $\lambda_{\rm eff}$ is needed to overcome the suppression. Thus, a large $v_3$ is preferred such that the $VV$ and $YY$ modes are forbidden by kinematics. However, for a large $v_3$,  the heavy quarks and leptons {\it alone} cannot produce enough production rate of diphoton pairs to explain the excess; for example, $\sigma(pp\to H_3\to \gamma\gamma) \sim 1{\rm~fb}$ for $v_3\sim 2~{\rm TeV}$ and $\beta=\pm \sqrt{3}$. We emphasize that one has to include the loop contributions of charged scalars to obtain a sizable  rate of diphoton productions.  Figure~\ref{fig:diphoton}(c) and (d) show that the $H_3$ scalar in the models of $\beta=\pm \sqrt{3}$ could explain the diphoton excess at a large range of parameter space. Unfortunately, the other two models of $\beta=\pm 1/\sqrt{3}$ cannot produce enough diphoton pairs in the entire parameter space; see Figs.~\ref{fig:diphoton}(a) and (b).

\begin{figure} 
\includegraphics[scale=0.35]{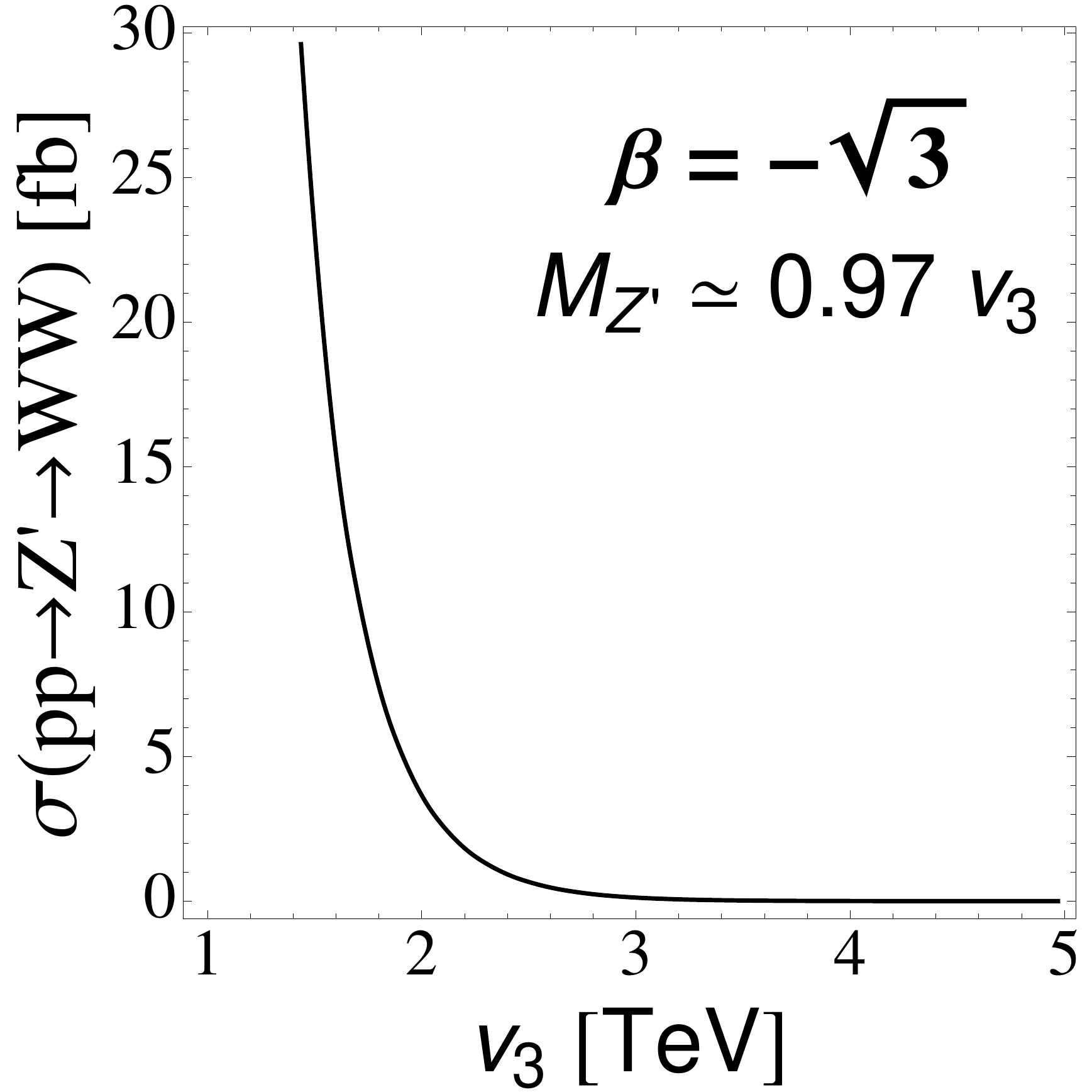}
\caption{\it The cross section of $pp\to Z^\prime \to W^+W^-$ as a function of $v_3$ in the model of $\beta=-\sqrt{3}$ at the 8 TeV LHC. }
\label{fig:zprime}
\end{figure}

Finally, we comment on the 2~TeV diboson excess observed at the LHC Run-I. The LHC Run-II data do  not confirm the excess, but the data do not significantly contradict it either. One has to await for more data to make affirmative conclusion.  The $Z^\prime$ boson is predominantly produced via the quark annihilation. It then could decay into a pair of $W$ bosons through the mixing with the $Z$ boson~\cite{Cao:2015lia}. Figure~\ref{fig:zprime} displays the cross section of $\sigma(pp\to Z^\prime\to W^+W^-)$ at the 8~TeV as a function of $v_3$ or $M_{Z^\prime}$($\simeq 0.97 v_3$) in the model of $\beta=-\sqrt{3}$. For a 2~TeV $Z^\prime$, the cross section of the diboson production is about 3~fb which is consistent with the LHC Run-I data.

\section{Conclusion}

In this work we explained both the 750~GeV diphoton and the 2~TeV diboson anomalies in the so-called 331 model which has the $SU(3)_C \otimes SU(3)_L \otimes U(1)_X$ gauge symmetry. The model consists of necessary new ingredients to explain both anomalies, e.g. new  heavy scalars, heavy quarks and leptons, and new gauge bosons. The symmetry breaking is induced by three scalar triplets. One of them, the scalar triplet $\chi$, is responsible for breaking  the $SU(3)_L\otimes U(1)_X$ gauge symmetry down to the SM electroweak symmetry $SU(2)_L\otimes U(1)_Y$ at the scale $v_3$.  The other two scalar triplets, $\rho$ and $\eta$, are in charge of the symmetry breaking of $SU(2)_L\otimes U(1)_Y \to U(1)_{\rm em}$ at the electroweak scale. The scalar $H_3$, arising from the $\chi$ scalar triplet, couples weakly to the SM particles at the tree level. It however can couple to the SM gluon and photons via the loops involving heavy quarks and leptons, charged scalars, and charged vector bosons. We use low energy theorem to calculate effective coupling of $H_3gg$, $H_3\gamma\gamma$, $H_3ZZ$, $H_3WW$ and $H_3Z\gamma$. The analytical results can be applied to new physics models satisfying the low energy theorem. We show  that $H_3$  is a good candidate of the 750~GeV diphoton resonance. We also demonstrate that the 2~TeV $Z^\prime$ boson can explain the diboson excess observed at the LHC Run-I. It would be interesting to see if future LHC data will confirm or rule out the two anomalies.

\begin{acknowledgements}
The work is supported in part by the National Science Foundation of China under Grand No. 11275009.
\end{acknowledgements}

\bibliographystyle{apsrev}
\bibliography{reference}

\begin{thebibliography}{121}
\expandafter\ifx\csname natexlab\endcsname\relax\def\natexlab#1{#1}\fi
\expandafter\ifx\csname bibnamefont\endcsname\relax
  \def\bibnamefont#1{#1}\fi
\expandafter\ifx\csname bibfnamefont\endcsname\relax
  \def\bibfnamefont#1{#1}\fi
\expandafter\ifx\csname citenamefont\endcsname\relax
  \def\citenamefont#1{#1}\fi
\expandafter\ifx\csname url\endcsname\relax
  \def\url#1{\texttt{#1}}\fi
\expandafter\ifx\csname urlprefix\endcsname\relax\def\urlprefix{URL }\fi
\providecommand{\bibinfo}[2]{#2}
\providecommand{\eprint}[2][]{\url{#2}}

\bibitem[{\citenamefont{${\rm The~ATLAS~Collaboration}$}(2015)}]{ATLAS-CONF-2015-081}
\bibinfo{author}{\bibfnamefont{${\rm The~ATLAS~Collaboration}$}}
  (\bibinfo{year}{2015}), \eprint{CMS-PAS-EXO-15-004}

\bibitem[{\citenamefont{${\rm The~CMS~Collaboration}$}(2015)}]{CMS:2015dxe}
\bibinfo{author}{\bibnamefont{${\rm The~CMS~Collaboration}$}}
  (\bibinfo{year}{2015}), \eprint{ATLAS-CONF-2015-081}.

\bibitem[{ATL(2016)}]{ATLAS-CONF-2016-018}
\bibinfo{type}{Tech. Rep.} \bibinfo{number}{ATLAS-CONF-2016-018},
  \bibinfo{institution}{CERN}, \bibinfo{address}{Geneva}
  (\bibinfo{year}{2016}), \urlprefix\url{http://cds.cern.ch/record/2141568}.

\bibitem[{CMS(2016)}]{CMS-PAS-EXO-16-018}
\bibinfo{type}{Tech. Rep.} \bibinfo{number}{CMS-PAS-EXO-16-018},
  \bibinfo{institution}{CERN}, \bibinfo{address}{Geneva}
  (\bibinfo{year}{2016}), \urlprefix\url{https://cds.cern.ch/record/2139899}.

\bibitem[{\citenamefont{Backovic et~al.}(2015)\citenamefont{Backovic, Mariotti,
  and Redigolo}}]{Backovic:2015fnp}
\bibinfo{author}{\bibfnamefont{M.}~\bibnamefont{Backovic}},
  \bibinfo{author}{\bibfnamefont{A.}~\bibnamefont{Mariotti}}, \bibnamefont{and}
  \bibinfo{author}{\bibfnamefont{D.}~\bibnamefont{Redigolo}}
  (\bibinfo{year}{2015}), \eprint{1512.04917}.

\bibitem[{\citenamefont{Harigaya and Nomura}(2015)}]{Harigaya:2015ezk}
\bibinfo{author}{\bibfnamefont{K.}~\bibnamefont{Harigaya}} \bibnamefont{and}
  \bibinfo{author}{\bibfnamefont{Y.}~\bibnamefont{Nomura}}
  (\bibinfo{year}{2015}), \eprint{1512.04850}.

\bibitem[{\citenamefont{Mambrini et~al.}(2015)\citenamefont{Mambrini, Arcadi,
  and Djouadi}}]{Mambrini:2015wyu}
\bibinfo{author}{\bibfnamefont{Y.}~\bibnamefont{Mambrini}},
  \bibinfo{author}{\bibfnamefont{G.}~\bibnamefont{Arcadi}}, \bibnamefont{and}
  \bibinfo{author}{\bibfnamefont{A.}~\bibnamefont{Djouadi}}
  (\bibinfo{year}{2015}), \eprint{1512.04913}.

\bibitem[{\citenamefont{Angelescu et~al.}(2015)\citenamefont{Angelescu,
  Djouadi, and Moreau}}]{Angelescu:2015uiz}
\bibinfo{author}{\bibfnamefont{A.}~\bibnamefont{Angelescu}},
  \bibinfo{author}{\bibfnamefont{A.}~\bibnamefont{Djouadi}}, \bibnamefont{and}
  \bibinfo{author}{\bibfnamefont{G.}~\bibnamefont{Moreau}}
  (\bibinfo{year}{2015}), \eprint{1512.04921}.

\bibitem[{\citenamefont{Pilaftsis}(2015)}]{Pilaftsis:2015ycr}
\bibinfo{author}{\bibfnamefont{A.}~\bibnamefont{Pilaftsis}}
  (\bibinfo{year}{2015}), \eprint{1512.04931}.

\bibitem[{\citenamefont{Franceschini et~al.}(2015)\citenamefont{Franceschini,
  Giudice, Kamenik, McCullough, Pomarol, Rattazzi, Redi, Riva, Strumia, and
  Torre}}]{Franceschini:2015kwy}
\bibinfo{author}{\bibfnamefont{R.}~\bibnamefont{Franceschini}},
  \bibinfo{author}{\bibfnamefont{G.~F.} \bibnamefont{Giudice}},
  \bibinfo{author}{\bibfnamefont{J.~F.} \bibnamefont{Kamenik}},
  \bibinfo{author}{\bibfnamefont{M.}~\bibnamefont{McCullough}},
  \bibinfo{author}{\bibfnamefont{A.}~\bibnamefont{Pomarol}},
  \bibinfo{author}{\bibfnamefont{R.}~\bibnamefont{Rattazzi}},
  \bibinfo{author}{\bibfnamefont{M.}~\bibnamefont{Redi}},
  \bibinfo{author}{\bibfnamefont{F.}~\bibnamefont{Riva}},
  \bibinfo{author}{\bibfnamefont{A.}~\bibnamefont{Strumia}}, \bibnamefont{and}
  \bibinfo{author}{\bibfnamefont{R.}~\bibnamefont{Torre}}
  (\bibinfo{year}{2015}), \eprint{1512.04933}.

\bibitem[{\citenamefont{Di~Chiara et~al.}(2015)\citenamefont{Di~Chiara,
  Marzola, and Raidal}}]{DiChiara:2015vdm}
\bibinfo{author}{\bibfnamefont{S.}~\bibnamefont{Di~Chiara}},
  \bibinfo{author}{\bibfnamefont{L.}~\bibnamefont{Marzola}}, \bibnamefont{and}
  \bibinfo{author}{\bibfnamefont{M.}~\bibnamefont{Raidal}}
  (\bibinfo{year}{2015}), \eprint{1512.04939}.

\bibitem[{\citenamefont{Low et~al.}(2015)\citenamefont{Low, Tesi, and
  Wang}}]{Low:2015qep}
\bibinfo{author}{\bibfnamefont{M.}~\bibnamefont{Low}},
  \bibinfo{author}{\bibfnamefont{A.}~\bibnamefont{Tesi}}, \bibnamefont{and}
  \bibinfo{author}{\bibfnamefont{L.-T.} \bibnamefont{Wang}}
  (\bibinfo{year}{2015}), \eprint{1512.05328}.

\bibitem[{\citenamefont{Bellazzini et~al.}(2015)\citenamefont{Bellazzini,
  Franceschini, Sala, and Serra}}]{Bellazzini:2015nxw}
\bibinfo{author}{\bibfnamefont{B.}~\bibnamefont{Bellazzini}},
  \bibinfo{author}{\bibfnamefont{R.}~\bibnamefont{Franceschini}},
  \bibinfo{author}{\bibfnamefont{F.}~\bibnamefont{Sala}}, \bibnamefont{and}
  \bibinfo{author}{\bibfnamefont{J.}~\bibnamefont{Serra}}
  (\bibinfo{year}{2015}), \eprint{1512.05330}.

\bibitem[{\citenamefont{Ellis et~al.}(2015)\citenamefont{Ellis, Ellis,
  Quevillon, Sanz, and You}}]{Ellis:2015oso}
\bibinfo{author}{\bibfnamefont{J.}~\bibnamefont{Ellis}},
  \bibinfo{author}{\bibfnamefont{S.~A.~R.} \bibnamefont{Ellis}},
  \bibinfo{author}{\bibfnamefont{J.}~\bibnamefont{Quevillon}},
  \bibinfo{author}{\bibfnamefont{V.}~\bibnamefont{Sanz}}, \bibnamefont{and}
  \bibinfo{author}{\bibfnamefont{T.}~\bibnamefont{You}} (\bibinfo{year}{2015}),
  \eprint{1512.05327}.

\bibitem[{\citenamefont{McDermott et~al.}(2015)\citenamefont{McDermott, Meade,
  and Ramani}}]{McDermott:2015sck}
\bibinfo{author}{\bibfnamefont{S.~D.} \bibnamefont{McDermott}},
  \bibinfo{author}{\bibfnamefont{P.}~\bibnamefont{Meade}}, \bibnamefont{and}
  \bibinfo{author}{\bibfnamefont{H.}~\bibnamefont{Ramani}}
  (\bibinfo{year}{2015}), \eprint{1512.05326}.

\bibitem[{\citenamefont{Higaki et~al.}(2015)\citenamefont{Higaki, Jeong,
  Kitajima, and Takahashi}}]{Higaki:2015jag}
\bibinfo{author}{\bibfnamefont{T.}~\bibnamefont{Higaki}},
  \bibinfo{author}{\bibfnamefont{K.~S.} \bibnamefont{Jeong}},
  \bibinfo{author}{\bibfnamefont{N.}~\bibnamefont{Kitajima}}, \bibnamefont{and}
  \bibinfo{author}{\bibfnamefont{F.}~\bibnamefont{Takahashi}}
  (\bibinfo{year}{2015}), \eprint{1512.05295}.

\bibitem[{\citenamefont{Gupta et~al.}(2015)\citenamefont{Gupta, Jäger, Kats,
  Perez, and Stamou}}]{Gupta:2015zzs}
\bibinfo{author}{\bibfnamefont{R.~S.} \bibnamefont{Gupta}},
  \bibinfo{author}{\bibfnamefont{S.}~\bibnamefont{Jäger}},
  \bibinfo{author}{\bibfnamefont{Y.}~\bibnamefont{Kats}},
  \bibinfo{author}{\bibfnamefont{G.}~\bibnamefont{Perez}}, \bibnamefont{and}
  \bibinfo{author}{\bibfnamefont{E.}~\bibnamefont{Stamou}}
  (\bibinfo{year}{2015}), \eprint{1512.05332}.

\bibitem[{\citenamefont{Petersson and Torre}(2015)}]{Petersson:2015mkr}
\bibinfo{author}{\bibfnamefont{C.}~\bibnamefont{Petersson}} \bibnamefont{and}
  \bibinfo{author}{\bibfnamefont{R.}~\bibnamefont{Torre}}
  (\bibinfo{year}{2015}), \eprint{1512.05333}.

\bibitem[{\citenamefont{Molinaro et~al.}(2015)\citenamefont{Molinaro, Sannino,
  and Vignaroli}}]{Molinaro:2015cwg}
\bibinfo{author}{\bibfnamefont{E.}~\bibnamefont{Molinaro}},
  \bibinfo{author}{\bibfnamefont{F.}~\bibnamefont{Sannino}}, \bibnamefont{and}
  \bibinfo{author}{\bibfnamefont{N.}~\bibnamefont{Vignaroli}}
  (\bibinfo{year}{2015}), \eprint{1512.05334}.

\bibitem[{\citenamefont{Nakai et~al.}(2015)\citenamefont{Nakai, Sato, and
  Tobioka}}]{Nakai:2015ptz}
\bibinfo{author}{\bibfnamefont{Y.}~\bibnamefont{Nakai}},
  \bibinfo{author}{\bibfnamefont{R.}~\bibnamefont{Sato}}, \bibnamefont{and}
  \bibinfo{author}{\bibfnamefont{K.}~\bibnamefont{Tobioka}}
  (\bibinfo{year}{2015}), \eprint{1512.04924}.

\bibitem[{\citenamefont{Buttazzo et~al.}(2015)\citenamefont{Buttazzo, Greljo,
  and Marzocca}}]{Buttazzo:2015txu}
\bibinfo{author}{\bibfnamefont{D.}~\bibnamefont{Buttazzo}},
  \bibinfo{author}{\bibfnamefont{A.}~\bibnamefont{Greljo}}, \bibnamefont{and}
  \bibinfo{author}{\bibfnamefont{D.}~\bibnamefont{Marzocca}}
  (\bibinfo{year}{2015}), \eprint{1512.04929}.

\bibitem[{\citenamefont{Bai et~al.}(2015)\citenamefont{Bai, Berger, and
  Lu}}]{Bai:2015nbs}
\bibinfo{author}{\bibfnamefont{Y.}~\bibnamefont{Bai}},
  \bibinfo{author}{\bibfnamefont{J.}~\bibnamefont{Berger}}, \bibnamefont{and}
  \bibinfo{author}{\bibfnamefont{R.}~\bibnamefont{Lu}} (\bibinfo{year}{2015}),
  \eprint{1512.05779}.

\bibitem[{\citenamefont{Aloni et~al.}(2015)\citenamefont{Aloni, Blum, Dery,
  Efrati, and Nir}}]{Aloni:2015mxa}
\bibinfo{author}{\bibfnamefont{D.}~\bibnamefont{Aloni}},
  \bibinfo{author}{\bibfnamefont{K.}~\bibnamefont{Blum}},
  \bibinfo{author}{\bibfnamefont{A.}~\bibnamefont{Dery}},
  \bibinfo{author}{\bibfnamefont{A.}~\bibnamefont{Efrati}}, \bibnamefont{and}
  \bibinfo{author}{\bibfnamefont{Y.}~\bibnamefont{Nir}} (\bibinfo{year}{2015}),
  \eprint{1512.05778}.

\bibitem[{\citenamefont{Falkowski et~al.}(2015)\citenamefont{Falkowski, Slone,
  and Volansky}}]{Falkowski:2015swt}
\bibinfo{author}{\bibfnamefont{A.}~\bibnamefont{Falkowski}},
  \bibinfo{author}{\bibfnamefont{O.}~\bibnamefont{Slone}}, \bibnamefont{and}
  \bibinfo{author}{\bibfnamefont{T.}~\bibnamefont{Volansky}}
  (\bibinfo{year}{2015}), \eprint{1512.05777}.

\bibitem[{\citenamefont{Csaki et~al.}(2015)\citenamefont{Csaki, Hubisz, and
  Terning}}]{Csaki:2015vek}
\bibinfo{author}{\bibfnamefont{C.}~\bibnamefont{Csaki}},
  \bibinfo{author}{\bibfnamefont{J.}~\bibnamefont{Hubisz}}, \bibnamefont{and}
  \bibinfo{author}{\bibfnamefont{J.}~\bibnamefont{Terning}}
  (\bibinfo{year}{2015}), \eprint{1512.05776}.

\bibitem[{\citenamefont{Agrawal et~al.}(2015)\citenamefont{Agrawal, Fan,
  Heidenreich, Reece, and Strassler}}]{Agrawal:2015dbf}
\bibinfo{author}{\bibfnamefont{P.}~\bibnamefont{Agrawal}},
  \bibinfo{author}{\bibfnamefont{J.}~\bibnamefont{Fan}},
  \bibinfo{author}{\bibfnamefont{B.}~\bibnamefont{Heidenreich}},
  \bibinfo{author}{\bibfnamefont{M.}~\bibnamefont{Reece}}, \bibnamefont{and}
  \bibinfo{author}{\bibfnamefont{M.}~\bibnamefont{Strassler}}
  (\bibinfo{year}{2015}), \eprint{1512.05775}.

\bibitem[{\citenamefont{Ahmed et~al.}(2015)\citenamefont{Ahmed, Dillon,
  Grzadkowski, Gunion, and Jiang}}]{Ahmed:2015uqt}
\bibinfo{author}{\bibfnamefont{A.}~\bibnamefont{Ahmed}},
  \bibinfo{author}{\bibfnamefont{B.~M.} \bibnamefont{Dillon}},
  \bibinfo{author}{\bibfnamefont{B.}~\bibnamefont{Grzadkowski}},
  \bibinfo{author}{\bibfnamefont{J.~F.} \bibnamefont{Gunion}},
  \bibnamefont{and} \bibinfo{author}{\bibfnamefont{Y.}~\bibnamefont{Jiang}}
  (\bibinfo{year}{2015}), \eprint{1512.05771}.

\bibitem[{\citenamefont{Chakrabortty et~al.}(2015)\citenamefont{Chakrabortty,
  Choudhury, Ghosh, Mondal, and Srivastava}}]{Chakrabortty:2015hff}
\bibinfo{author}{\bibfnamefont{J.}~\bibnamefont{Chakrabortty}},
  \bibinfo{author}{\bibfnamefont{A.}~\bibnamefont{Choudhury}},
  \bibinfo{author}{\bibfnamefont{P.}~\bibnamefont{Ghosh}},
  \bibinfo{author}{\bibfnamefont{S.}~\bibnamefont{Mondal}}, \bibnamefont{and}
  \bibinfo{author}{\bibfnamefont{T.}~\bibnamefont{Srivastava}}
  (\bibinfo{year}{2015}), \eprint{1512.05767}.

\bibitem[{\citenamefont{Bian et~al.}(2015)\citenamefont{Bian, Chen, Liu, and
  Shu}}]{Bian:2015kjt}
\bibinfo{author}{\bibfnamefont{L.}~\bibnamefont{Bian}},
  \bibinfo{author}{\bibfnamefont{N.}~\bibnamefont{Chen}},
  \bibinfo{author}{\bibfnamefont{D.}~\bibnamefont{Liu}}, \bibnamefont{and}
  \bibinfo{author}{\bibfnamefont{J.}~\bibnamefont{Shu}} (\bibinfo{year}{2015}),
  \eprint{1512.05759}.

\bibitem[{\citenamefont{Curtin and Verhaaren}(2015)}]{Curtin:2015jcv}
\bibinfo{author}{\bibfnamefont{D.}~\bibnamefont{Curtin}} \bibnamefont{and}
  \bibinfo{author}{\bibfnamefont{C.~B.} \bibnamefont{Verhaaren}}
  (\bibinfo{year}{2015}), \eprint{1512.05753}.

\bibitem[{\citenamefont{Fichet et~al.}(2015)\citenamefont{Fichet, von
  Gersdorff, and Royon}}]{Fichet:2015vvy}
\bibinfo{author}{\bibfnamefont{S.}~\bibnamefont{Fichet}},
  \bibinfo{author}{\bibfnamefont{G.}~\bibnamefont{von Gersdorff}},
  \bibnamefont{and} \bibinfo{author}{\bibfnamefont{C.}~\bibnamefont{Royon}}
  (\bibinfo{year}{2015}), \eprint{1512.05751}.

\bibitem[{\citenamefont{Chao et~al.}(2015)\citenamefont{Chao, Huo, and
  Yu}}]{Chao:2015ttq}
\bibinfo{author}{\bibfnamefont{W.}~\bibnamefont{Chao}},
  \bibinfo{author}{\bibfnamefont{R.}~\bibnamefont{Huo}}, \bibnamefont{and}
  \bibinfo{author}{\bibfnamefont{J.-H.} \bibnamefont{Yu}}
  (\bibinfo{year}{2015}), \eprint{1512.05738}.

\bibitem[{\citenamefont{Demidov and Gorbunov}(2015)}]{Demidov:2015zqn}
\bibinfo{author}{\bibfnamefont{S.~V.} \bibnamefont{Demidov}} \bibnamefont{and}
  \bibinfo{author}{\bibfnamefont{D.~S.} \bibnamefont{Gorbunov}}
  (\bibinfo{year}{2015}), \eprint{1512.05723}.

\bibitem[{\citenamefont{No et~al.}(2015)\citenamefont{No, Sanz, and
  Setford}}]{No:2015bsn}
\bibinfo{author}{\bibfnamefont{J.~M.} \bibnamefont{No}},
  \bibinfo{author}{\bibfnamefont{V.}~\bibnamefont{Sanz}}, \bibnamefont{and}
  \bibinfo{author}{\bibfnamefont{J.}~\bibnamefont{Setford}}
  (\bibinfo{year}{2015}), \eprint{1512.05700}.

\bibitem[{\citenamefont{Becirevic et~al.}(2015)\citenamefont{Becirevic,
  Bertuzzo, Sumensari, and Funchal}}]{Becirevic:2015fmu}
\bibinfo{author}{\bibfnamefont{D.}~\bibnamefont{Becirevic}},
  \bibinfo{author}{\bibfnamefont{E.}~\bibnamefont{Bertuzzo}},
  \bibinfo{author}{\bibfnamefont{O.}~\bibnamefont{Sumensari}},
  \bibnamefont{and} \bibinfo{author}{\bibfnamefont{R.~Z.}
  \bibnamefont{Funchal}} (\bibinfo{year}{2015}), \eprint{1512.05623}.

\bibitem[{\citenamefont{Cox et~al.}(2015)\citenamefont{Cox, Medina, Ray, and
  Spray}}]{Cox:2015ckc}
\bibinfo{author}{\bibfnamefont{P.}~\bibnamefont{Cox}},
  \bibinfo{author}{\bibfnamefont{A.~D.} \bibnamefont{Medina}},
  \bibinfo{author}{\bibfnamefont{T.~S.} \bibnamefont{Ray}}, \bibnamefont{and}
  \bibinfo{author}{\bibfnamefont{A.}~\bibnamefont{Spray}}
  (\bibinfo{year}{2015}), \eprint{1512.05618}.

\bibitem[{\citenamefont{Kobakhidze et~al.}(2015)\citenamefont{Kobakhidze, Wang,
  Wu, Yang, and Zhang}}]{Kobakhidze:2015ldh}
\bibinfo{author}{\bibfnamefont{A.}~\bibnamefont{Kobakhidze}},
  \bibinfo{author}{\bibfnamefont{F.}~\bibnamefont{Wang}},
  \bibinfo{author}{\bibfnamefont{L.}~\bibnamefont{Wu}},
  \bibinfo{author}{\bibfnamefont{J.~M.} \bibnamefont{Yang}}, \bibnamefont{and}
  \bibinfo{author}{\bibfnamefont{M.}~\bibnamefont{Zhang}}
  (\bibinfo{year}{2015}), \eprint{1512.05585}.

\bibitem[{\citenamefont{Matsuzaki and Yamawaki}(2015)}]{Matsuzaki:2015che}
\bibinfo{author}{\bibfnamefont{S.}~\bibnamefont{Matsuzaki}} \bibnamefont{and}
  \bibinfo{author}{\bibfnamefont{K.}~\bibnamefont{Yamawaki}}
  (\bibinfo{year}{2015}), \eprint{1512.05564}.

\bibitem[{\citenamefont{Cao et~al.}(2015{\natexlab{a}})\citenamefont{Cao, Liu,
  Xie, Yan, and Zhang}}]{Cao:2015pto}
\bibinfo{author}{\bibfnamefont{Q.-H.} \bibnamefont{Cao}},
  \bibinfo{author}{\bibfnamefont{Y.}~\bibnamefont{Liu}},
  \bibinfo{author}{\bibfnamefont{K.-P.} \bibnamefont{Xie}},
  \bibinfo{author}{\bibfnamefont{B.}~\bibnamefont{Yan}}, \bibnamefont{and}
  \bibinfo{author}{\bibfnamefont{D.-M.} \bibnamefont{Zhang}}
  (\bibinfo{year}{2015}{\natexlab{a}}), \eprint{1512.05542}.

\bibitem[{\citenamefont{Dutta et~al.}(2015)\citenamefont{Dutta, Gao, Ghosh,
  Gogoladze, and Li}}]{Dutta:2015wqh}
\bibinfo{author}{\bibfnamefont{B.}~\bibnamefont{Dutta}},
  \bibinfo{author}{\bibfnamefont{Y.}~\bibnamefont{Gao}},
  \bibinfo{author}{\bibfnamefont{T.}~\bibnamefont{Ghosh}},
  \bibinfo{author}{\bibfnamefont{I.}~\bibnamefont{Gogoladze}},
  \bibnamefont{and} \bibinfo{author}{\bibfnamefont{T.}~\bibnamefont{Li}}
  (\bibinfo{year}{2015}), \eprint{1512.05439}.

\bibitem[{\citenamefont{Benbrik et~al.}(2015)\citenamefont{Benbrik, Chen, and
  Nomura}}]{Benbrik:2015fyz}
\bibinfo{author}{\bibfnamefont{R.}~\bibnamefont{Benbrik}},
  \bibinfo{author}{\bibfnamefont{C.-H.} \bibnamefont{Chen}}, \bibnamefont{and}
  \bibinfo{author}{\bibfnamefont{T.}~\bibnamefont{Nomura}}
  (\bibinfo{year}{2015}), \eprint{1512.06028}.

\bibitem[{\citenamefont{Megias et~al.}(2015)\citenamefont{Megias, Pujolas, and
  Quiros}}]{Megias:2015ory}
\bibinfo{author}{\bibfnamefont{E.}~\bibnamefont{Megias}},
  \bibinfo{author}{\bibfnamefont{O.}~\bibnamefont{Pujolas}}, \bibnamefont{and}
  \bibinfo{author}{\bibfnamefont{M.}~\bibnamefont{Quiros}}
  (\bibinfo{year}{2015}), \eprint{1512.06106}.

\bibitem[{\citenamefont{Carpenter et~al.}(2015)\citenamefont{Carpenter,
  Colburn, and Goodman}}]{Carpenter:2015ucu}
\bibinfo{author}{\bibfnamefont{L.~M.} \bibnamefont{Carpenter}},
  \bibinfo{author}{\bibfnamefont{R.}~\bibnamefont{Colburn}}, \bibnamefont{and}
  \bibinfo{author}{\bibfnamefont{J.}~\bibnamefont{Goodman}}
  (\bibinfo{year}{2015}), \eprint{1512.06107}.

\bibitem[{\citenamefont{Bernon and Smith}(2015)}]{Bernon:2015abk}
\bibinfo{author}{\bibfnamefont{J.}~\bibnamefont{Bernon}} \bibnamefont{and}
  \bibinfo{author}{\bibfnamefont{C.}~\bibnamefont{Smith}}
  (\bibinfo{year}{2015}), \eprint{1512.06113}.

\bibitem[{\citenamefont{Alves et~al.}(2015)\citenamefont{Alves, Dias, and
  Sinha}}]{Alves:2015jgx}
\bibinfo{author}{\bibfnamefont{A.}~\bibnamefont{Alves}},
  \bibinfo{author}{\bibfnamefont{A.~G.} \bibnamefont{Dias}}, \bibnamefont{and}
  \bibinfo{author}{\bibfnamefont{K.}~\bibnamefont{Sinha}}
  (\bibinfo{year}{2015}), \eprint{1512.06091}.

\bibitem[{\citenamefont{Gabrielli et~al.}(2015)\citenamefont{Gabrielli,
  Kannike, Mele, Raidal, Spethmann, and Veermäe}}]{Gabrielli:2015dhk}
\bibinfo{author}{\bibfnamefont{E.}~\bibnamefont{Gabrielli}},
  \bibinfo{author}{\bibfnamefont{K.}~\bibnamefont{Kannike}},
  \bibinfo{author}{\bibfnamefont{B.}~\bibnamefont{Mele}},
  \bibinfo{author}{\bibfnamefont{M.}~\bibnamefont{Raidal}},
  \bibinfo{author}{\bibfnamefont{C.}~\bibnamefont{Spethmann}},
  \bibnamefont{and} \bibinfo{author}{\bibfnamefont{H.}~\bibnamefont{Veermäe}}
  (\bibinfo{year}{2015}), \eprint{1512.05961}.

\bibitem[{\citenamefont{Chao}(2015)}]{Chao:2015nsm}
\bibinfo{author}{\bibfnamefont{W.}~\bibnamefont{Chao}} (\bibinfo{year}{2015}),
  \eprint{1512.06297}.

\bibitem[{\citenamefont{Arun and Saha}(2015)}]{Arun:2015ubr}
\bibinfo{author}{\bibfnamefont{M.~T.} \bibnamefont{Arun}} \bibnamefont{and}
  \bibinfo{author}{\bibfnamefont{P.}~\bibnamefont{Saha}}
  (\bibinfo{year}{2015}), \eprint{1512.06335}.

\bibitem[{\citenamefont{Han et~al.}(2015{\natexlab{a}})\citenamefont{Han, Lee,
  Park, and Sanz}}]{Han:2015cty}
\bibinfo{author}{\bibfnamefont{C.}~\bibnamefont{Han}},
  \bibinfo{author}{\bibfnamefont{H.~M.} \bibnamefont{Lee}},
  \bibinfo{author}{\bibfnamefont{M.}~\bibnamefont{Park}}, \bibnamefont{and}
  \bibinfo{author}{\bibfnamefont{V.}~\bibnamefont{Sanz}}
  (\bibinfo{year}{2015}{\natexlab{a}}), \eprint{1512.06376}.

\bibitem[{\citenamefont{Chang}(2015)}]{Chang:2015bzc}
\bibinfo{author}{\bibfnamefont{S.}~\bibnamefont{Chang}} (\bibinfo{year}{2015}),
  \eprint{1512.06426}.

\bibitem[{\citenamefont{Chakraborty and Kundu}(2015)}]{Chakraborty:2015jvs}
\bibinfo{author}{\bibfnamefont{I.}~\bibnamefont{Chakraborty}} \bibnamefont{and}
  \bibinfo{author}{\bibfnamefont{A.}~\bibnamefont{Kundu}}
  (\bibinfo{year}{2015}), \eprint{1512.06508}.

\bibitem[{\citenamefont{Ding et~al.}(2015)\citenamefont{Ding, Huang, Li, and
  Zhu}}]{Ding:2015rxx}
\bibinfo{author}{\bibfnamefont{R.}~\bibnamefont{Ding}},
  \bibinfo{author}{\bibfnamefont{L.}~\bibnamefont{Huang}},
  \bibinfo{author}{\bibfnamefont{T.}~\bibnamefont{Li}}, \bibnamefont{and}
  \bibinfo{author}{\bibfnamefont{B.}~\bibnamefont{Zhu}} (\bibinfo{year}{2015}),
  \eprint{1512.06560}.

\bibitem[{\citenamefont{Han et~al.}(2015{\natexlab{b}})\citenamefont{Han, Wang,
  and Zheng}}]{Han:2015dlp}
\bibinfo{author}{\bibfnamefont{H.}~\bibnamefont{Han}},
  \bibinfo{author}{\bibfnamefont{S.}~\bibnamefont{Wang}}, \bibnamefont{and}
  \bibinfo{author}{\bibfnamefont{S.}~\bibnamefont{Zheng}}
  (\bibinfo{year}{2015}{\natexlab{b}}), \eprint{1512.06562}.

\bibitem[{\citenamefont{Han and Wang}(2015)}]{Han:2015qqj}
\bibinfo{author}{\bibfnamefont{X.-F.} \bibnamefont{Han}} \bibnamefont{and}
  \bibinfo{author}{\bibfnamefont{L.}~\bibnamefont{Wang}}
  (\bibinfo{year}{2015}), \eprint{1512.06587}.

\bibitem[{\citenamefont{Luo et~al.}(2015)\citenamefont{Luo, Wang, Xu, Zhang,
  and Zhu}}]{Luo:2015yio}
\bibinfo{author}{\bibfnamefont{M.-x.} \bibnamefont{Luo}},
  \bibinfo{author}{\bibfnamefont{K.}~\bibnamefont{Wang}},
  \bibinfo{author}{\bibfnamefont{T.}~\bibnamefont{Xu}},
  \bibinfo{author}{\bibfnamefont{L.}~\bibnamefont{Zhang}}, \bibnamefont{and}
  \bibinfo{author}{\bibfnamefont{G.}~\bibnamefont{Zhu}} (\bibinfo{year}{2015}),
  \eprint{1512.06670}.

\bibitem[{\citenamefont{Chang et~al.}(2015)\citenamefont{Chang, Cheung, and
  Lu}}]{Chang:2015sdy}
\bibinfo{author}{\bibfnamefont{J.}~\bibnamefont{Chang}},
  \bibinfo{author}{\bibfnamefont{K.}~\bibnamefont{Cheung}}, \bibnamefont{and}
  \bibinfo{author}{\bibfnamefont{C.-T.} \bibnamefont{Lu}}
  (\bibinfo{year}{2015}), \eprint{1512.06671}.

\bibitem[{\citenamefont{Bardhan et~al.}(2015)\citenamefont{Bardhan, Bhatia,
  Chakraborty, Maitra, Raychaudhuri, and Samui}}]{Bardhan:2015hcr}
\bibinfo{author}{\bibfnamefont{D.}~\bibnamefont{Bardhan}},
  \bibinfo{author}{\bibfnamefont{D.}~\bibnamefont{Bhatia}},
  \bibinfo{author}{\bibfnamefont{A.}~\bibnamefont{Chakraborty}},
  \bibinfo{author}{\bibfnamefont{U.}~\bibnamefont{Maitra}},
  \bibinfo{author}{\bibfnamefont{S.}~\bibnamefont{Raychaudhuri}},
  \bibnamefont{and} \bibinfo{author}{\bibfnamefont{T.}~\bibnamefont{Samui}}
  (\bibinfo{year}{2015}), \eprint{1512.06674}.

\bibitem[{\citenamefont{Feng et~al.}(2015)\citenamefont{Feng, Li, Zhang, and
  Zhao}}]{Feng:2015wil}
\bibinfo{author}{\bibfnamefont{T.-F.} \bibnamefont{Feng}},
  \bibinfo{author}{\bibfnamefont{X.-Q.} \bibnamefont{Li}},
  \bibinfo{author}{\bibfnamefont{H.-B.} \bibnamefont{Zhang}}, \bibnamefont{and}
  \bibinfo{author}{\bibfnamefont{S.-M.} \bibnamefont{Zhao}}
  (\bibinfo{year}{2015}), \eprint{1512.06696}.

\bibitem[{\citenamefont{Antipin et~al.}(2015)\citenamefont{Antipin, Mojaza, and
  Sannino}}]{Antipin:2015kgh}
\bibinfo{author}{\bibfnamefont{O.}~\bibnamefont{Antipin}},
  \bibinfo{author}{\bibfnamefont{M.}~\bibnamefont{Mojaza}}, \bibnamefont{and}
  \bibinfo{author}{\bibfnamefont{F.}~\bibnamefont{Sannino}}
  (\bibinfo{year}{2015}), \eprint{1512.06708}.

\bibitem[{\citenamefont{Wang et~al.}(2015)\citenamefont{Wang, Wu, Yang, and
  Zhang}}]{Wang:2015kuj}
\bibinfo{author}{\bibfnamefont{F.}~\bibnamefont{Wang}},
  \bibinfo{author}{\bibfnamefont{L.}~\bibnamefont{Wu}},
  \bibinfo{author}{\bibfnamefont{J.~M.} \bibnamefont{Yang}}, \bibnamefont{and}
  \bibinfo{author}{\bibfnamefont{M.}~\bibnamefont{Zhang}}
  (\bibinfo{year}{2015}), \eprint{1512.06715}.

\bibitem[{\citenamefont{Cao et~al.}(2015{\natexlab{b}})\citenamefont{Cao, Han,
  Shang, Su, Yang, and Zhang}}]{Cao:2015twy}
\bibinfo{author}{\bibfnamefont{J.}~\bibnamefont{Cao}},
  \bibinfo{author}{\bibfnamefont{C.}~\bibnamefont{Han}},
  \bibinfo{author}{\bibfnamefont{L.}~\bibnamefont{Shang}},
  \bibinfo{author}{\bibfnamefont{W.}~\bibnamefont{Su}},
  \bibinfo{author}{\bibfnamefont{J.~M.} \bibnamefont{Yang}}, \bibnamefont{and}
  \bibinfo{author}{\bibfnamefont{Y.}~\bibnamefont{Zhang}}
  (\bibinfo{year}{2015}{\natexlab{b}}), \eprint{1512.06728}.

\bibitem[{\citenamefont{Huang et~al.}(2015{\natexlab{a}})\citenamefont{Huang,
  Li, Liu, and Wang}}]{Huang:2015evq}
\bibinfo{author}{\bibfnamefont{F.~P.} \bibnamefont{Huang}},
  \bibinfo{author}{\bibfnamefont{C.~S.} \bibnamefont{Li}},
  \bibinfo{author}{\bibfnamefont{Z.~L.} \bibnamefont{Liu}}, \bibnamefont{and}
  \bibinfo{author}{\bibfnamefont{Y.}~\bibnamefont{Wang}}
  (\bibinfo{year}{2015}{\natexlab{a}}), \eprint{1512.06732}.

\bibitem[{\citenamefont{Liao and Zheng}(2015)}]{Liao:2015tow}
\bibinfo{author}{\bibfnamefont{W.}~\bibnamefont{Liao}} \bibnamefont{and}
  \bibinfo{author}{\bibfnamefont{H.-q.} \bibnamefont{Zheng}}
  (\bibinfo{year}{2015}), \eprint{1512.06741}.

\bibitem[{\citenamefont{Heckman}(2015)}]{Heckman:2015kqk}
\bibinfo{author}{\bibfnamefont{J.~J.} \bibnamefont{Heckman}}
  (\bibinfo{year}{2015}), \eprint{1512.06773}.

\bibitem[{\citenamefont{Bi et~al.}(2015)\citenamefont{Bi, Xiang, Yin, and
  Yu}}]{Bi:2015uqd}
\bibinfo{author}{\bibfnamefont{X.-J.} \bibnamefont{Bi}},
  \bibinfo{author}{\bibfnamefont{Q.-F.} \bibnamefont{Xiang}},
  \bibinfo{author}{\bibfnamefont{P.-F.} \bibnamefont{Yin}}, \bibnamefont{and}
  \bibinfo{author}{\bibfnamefont{Z.-H.} \bibnamefont{Yu}}
  (\bibinfo{year}{2015}), \eprint{1512.06787}.

\bibitem[{\citenamefont{Cho et~al.}(2015)\citenamefont{Cho, Kim, Kong, Lim,
  Matchev, Park, and Park}}]{Cho:2015nxy}
\bibinfo{author}{\bibfnamefont{W.~S.} \bibnamefont{Cho}},
  \bibinfo{author}{\bibfnamefont{D.}~\bibnamefont{Kim}},
  \bibinfo{author}{\bibfnamefont{K.}~\bibnamefont{Kong}},
  \bibinfo{author}{\bibfnamefont{S.~H.} \bibnamefont{Lim}},
  \bibinfo{author}{\bibfnamefont{K.~T.} \bibnamefont{Matchev}},
  \bibinfo{author}{\bibfnamefont{J.-C.} \bibnamefont{Park}}, \bibnamefont{and}
  \bibinfo{author}{\bibfnamefont{M.}~\bibnamefont{Park}}
  (\bibinfo{year}{2015}), \eprint{1512.06824}.

\bibitem[{\citenamefont{Cline and Liu}(2015)}]{Cline:2015msi}
\bibinfo{author}{\bibfnamefont{J.~M.} \bibnamefont{Cline}} \bibnamefont{and}
  \bibinfo{author}{\bibfnamefont{Z.}~\bibnamefont{Liu}} (\bibinfo{year}{2015}),
  \eprint{1512.06827}.

\bibitem[{\citenamefont{Bauer and Neubert}(2015)}]{Bauer:2015boy}
\bibinfo{author}{\bibfnamefont{M.}~\bibnamefont{Bauer}} \bibnamefont{and}
  \bibinfo{author}{\bibfnamefont{M.}~\bibnamefont{Neubert}}
  (\bibinfo{year}{2015}), \eprint{1512.06828}.

\bibitem[{\citenamefont{Barducci et~al.}(2015)\citenamefont{Barducci, Goudelis,
  Kulkarni, and Sengupta}}]{Barducci:2015gtd}
\bibinfo{author}{\bibfnamefont{D.}~\bibnamefont{Barducci}},
  \bibinfo{author}{\bibfnamefont{A.}~\bibnamefont{Goudelis}},
  \bibinfo{author}{\bibfnamefont{S.}~\bibnamefont{Kulkarni}}, \bibnamefont{and}
  \bibinfo{author}{\bibfnamefont{D.}~\bibnamefont{Sengupta}}
  (\bibinfo{year}{2015}), \eprint{1512.06842}.

\bibitem[{\citenamefont{Boucenna et~al.}(2015)\citenamefont{Boucenna, Morisi,
  and Vicente}}]{Boucenna:2015pav}
\bibinfo{author}{\bibfnamefont{S.~M.} \bibnamefont{Boucenna}},
  \bibinfo{author}{\bibfnamefont{S.}~\bibnamefont{Morisi}}, \bibnamefont{and}
  \bibinfo{author}{\bibfnamefont{A.}~\bibnamefont{Vicente}}
  (\bibinfo{year}{2015}), \eprint{1512.06878}.

\bibitem[{\citenamefont{Murphy}(2015)}]{Murphy:2015kag}
\bibinfo{author}{\bibfnamefont{C.~W.} \bibnamefont{Murphy}}
  (\bibinfo{year}{2015}), \eprint{1512.06976}.

\bibitem[{\citenamefont{Hernández and Nisandzic}(2015)}]{Hernandez:2015ywg}
\bibinfo{author}{\bibfnamefont{A.~E.~C.} \bibnamefont{Hernández}}
  \bibnamefont{and} \bibinfo{author}{\bibfnamefont{I.}~\bibnamefont{Nisandzic}}
  (\bibinfo{year}{2015}), \eprint{1512.07165}.

\bibitem[{\citenamefont{Dey et~al.}(2015)\citenamefont{Dey, Mohanty, and
  Tomar}}]{Dey:2015bur}
\bibinfo{author}{\bibfnamefont{U.~K.} \bibnamefont{Dey}},
  \bibinfo{author}{\bibfnamefont{S.}~\bibnamefont{Mohanty}}, \bibnamefont{and}
  \bibinfo{author}{\bibfnamefont{G.}~\bibnamefont{Tomar}}
  (\bibinfo{year}{2015}), \eprint{1512.07212}.

\bibitem[{\citenamefont{Pelaggi et~al.}(2015)\citenamefont{Pelaggi, Strumia,
  and Vigiani}}]{Pelaggi:2015knk}
\bibinfo{author}{\bibfnamefont{G.~M.} \bibnamefont{Pelaggi}},
  \bibinfo{author}{\bibfnamefont{A.}~\bibnamefont{Strumia}}, \bibnamefont{and}
  \bibinfo{author}{\bibfnamefont{E.}~\bibnamefont{Vigiani}}
  (\bibinfo{year}{2015}), \eprint{1512.07225}.

\bibitem[{\citenamefont{de~Blas et~al.}(2015)\citenamefont{de~Blas, Santiago,
  and Vega-Morales}}]{deBlas:2015hlv}
\bibinfo{author}{\bibfnamefont{J.}~\bibnamefont{de~Blas}},
  \bibinfo{author}{\bibfnamefont{J.}~\bibnamefont{Santiago}}, \bibnamefont{and}
  \bibinfo{author}{\bibfnamefont{R.}~\bibnamefont{Vega-Morales}}
  (\bibinfo{year}{2015}), \eprint{1512.07229}.

\bibitem[{\citenamefont{Belyaev et~al.}(2015)\citenamefont{Belyaev,
  Cacciapaglia, Cai, Flacke, Parolini, and Serôdio}}]{Belyaev:2015hgo}
\bibinfo{author}{\bibfnamefont{A.}~\bibnamefont{Belyaev}},
  \bibinfo{author}{\bibfnamefont{G.}~\bibnamefont{Cacciapaglia}},
  \bibinfo{author}{\bibfnamefont{H.}~\bibnamefont{Cai}},
  \bibinfo{author}{\bibfnamefont{T.}~\bibnamefont{Flacke}},
  \bibinfo{author}{\bibfnamefont{A.}~\bibnamefont{Parolini}}, \bibnamefont{and}
  \bibinfo{author}{\bibfnamefont{H.}~\bibnamefont{Serôdio}}
  (\bibinfo{year}{2015}), \eprint{1512.07242}.

\bibitem[{\citenamefont{Dev and Teresi}(2015)}]{Dev:2015isx}
\bibinfo{author}{\bibfnamefont{P.~S.~B.} \bibnamefont{Dev}} \bibnamefont{and}
  \bibinfo{author}{\bibfnamefont{D.}~\bibnamefont{Teresi}}
  (\bibinfo{year}{2015}), \eprint{1512.07243}.

\bibitem[{\citenamefont{Gu and Liu}(2015)}]{Gu:2015lxj}
\bibinfo{author}{\bibfnamefont{J.}~\bibnamefont{Gu}} \bibnamefont{and}
  \bibinfo{author}{\bibfnamefont{Z.}~\bibnamefont{Liu}} (\bibinfo{year}{2015}),
  \eprint{1512.07624}.

\bibitem[{\citenamefont{Cvetič et~al.}(2015)\citenamefont{Cvetič, Halverson,
  and Langacker}}]{Cvetic:2015vit}
\bibinfo{author}{\bibfnamefont{M.}~\bibnamefont{Cvetič}},
  \bibinfo{author}{\bibfnamefont{J.}~\bibnamefont{Halverson}},
  \bibnamefont{and} \bibinfo{author}{\bibfnamefont{P.}~\bibnamefont{Langacker}}
  (\bibinfo{year}{2015}), \eprint{1512.07622}.

\bibitem[{\citenamefont{Altmannshofer et~al.}(2015)\citenamefont{Altmannshofer,
  Galloway, Gori, Kagan, Martin, and Zupan}}]{Altmannshofer:2015xfo}
\bibinfo{author}{\bibfnamefont{W.}~\bibnamefont{Altmannshofer}},
  \bibinfo{author}{\bibfnamefont{J.}~\bibnamefont{Galloway}},
  \bibinfo{author}{\bibfnamefont{S.}~\bibnamefont{Gori}},
  \bibinfo{author}{\bibfnamefont{A.~L.} \bibnamefont{Kagan}},
  \bibinfo{author}{\bibfnamefont{A.}~\bibnamefont{Martin}}, \bibnamefont{and}
  \bibinfo{author}{\bibfnamefont{J.}~\bibnamefont{Zupan}}
  (\bibinfo{year}{2015}), \eprint{1512.07616}.

\bibitem[{\citenamefont{Cao et~al.}(2015{\natexlab{c}})\citenamefont{Cao, Chen,
  and Gu}}]{Cao:2015xjz}
\bibinfo{author}{\bibfnamefont{Q.-H.} \bibnamefont{Cao}},
  \bibinfo{author}{\bibfnamefont{S.-L.} \bibnamefont{Chen}}, \bibnamefont{and}
  \bibinfo{author}{\bibfnamefont{P.-H.} \bibnamefont{Gu}}
  (\bibinfo{year}{2015}{\natexlab{c}}), \eprint{1512.07541}.

\bibitem[{\citenamefont{Chakraborty et~al.}(2015)\citenamefont{Chakraborty,
  Chakraborty, and Raychaudhuri}}]{Chakraborty:2015gyj}
\bibinfo{author}{\bibfnamefont{S.}~\bibnamefont{Chakraborty}},
  \bibinfo{author}{\bibfnamefont{A.}~\bibnamefont{Chakraborty}},
  \bibnamefont{and}
  \bibinfo{author}{\bibfnamefont{S.}~\bibnamefont{Raychaudhuri}}
  (\bibinfo{year}{2015}), \eprint{1512.07527}.

\bibitem[{\citenamefont{Badziak}(2015)}]{Badziak:2015zez}
\bibinfo{author}{\bibfnamefont{M.}~\bibnamefont{Badziak}}
  (\bibinfo{year}{2015}), \eprint{1512.07497}.

\bibitem[{\citenamefont{Patel and Sharma}(2015)}]{Patel:2015ulo}
\bibinfo{author}{\bibfnamefont{K.~M.} \bibnamefont{Patel}} \bibnamefont{and}
  \bibinfo{author}{\bibfnamefont{P.}~\bibnamefont{Sharma}}
  (\bibinfo{year}{2015}), \eprint{1512.07468}.

\bibitem[{\citenamefont{Moretti and Yagyu}(2015)}]{Moretti:2015pbj}
\bibinfo{author}{\bibfnamefont{S.}~\bibnamefont{Moretti}} \bibnamefont{and}
  \bibinfo{author}{\bibfnamefont{K.}~\bibnamefont{Yagyu}}
  (\bibinfo{year}{2015}), \eprint{1512.07462}.

\bibitem[{\citenamefont{Huang et~al.}(2015{\natexlab{b}})\citenamefont{Huang,
  Tsai, and Yuan}}]{Huang:2015rkj}
\bibinfo{author}{\bibfnamefont{W.-C.} \bibnamefont{Huang}},
  \bibinfo{author}{\bibfnamefont{Y.-L.~S.} \bibnamefont{Tsai}},
  \bibnamefont{and} \bibinfo{author}{\bibfnamefont{T.-C.} \bibnamefont{Yuan}}
  (\bibinfo{year}{2015}{\natexlab{b}}), \eprint{1512.07268}.

\bibitem[{\citenamefont{Hall et~al.}(2015)\citenamefont{Hall, Harigaya, and
  Nomura}}]{Hall:2015xds}
\bibinfo{author}{\bibfnamefont{L.~J.} \bibnamefont{Hall}},
  \bibinfo{author}{\bibfnamefont{K.}~\bibnamefont{Harigaya}}, \bibnamefont{and}
  \bibinfo{author}{\bibfnamefont{Y.}~\bibnamefont{Nomura}}
  (\bibinfo{year}{2015}), \eprint{1512.07904}.

\bibitem[{\citenamefont{Casas et~al.}(2015)\citenamefont{Casas, Espinosa, and
  Moreno}}]{Casas:2015blx}
\bibinfo{author}{\bibfnamefont{J.~A.} \bibnamefont{Casas}},
  \bibinfo{author}{\bibfnamefont{J.~R.} \bibnamefont{Espinosa}},
  \bibnamefont{and} \bibinfo{author}{\bibfnamefont{J.~M.} \bibnamefont{Moreno}}
  (\bibinfo{year}{2015}), \eprint{1512.07895}.

\bibitem[{\citenamefont{Zhang and Zhou}(2015)}]{Zhang:2015uuo}
\bibinfo{author}{\bibfnamefont{J.}~\bibnamefont{Zhang}} \bibnamefont{and}
  \bibinfo{author}{\bibfnamefont{S.}~\bibnamefont{Zhou}}
  (\bibinfo{year}{2015}), \eprint{1512.07889}.

\bibitem[{\citenamefont{Liu et~al.}(2015)\citenamefont{Liu, Wang, and
  Xue}}]{Liu:2015yec}
\bibinfo{author}{\bibfnamefont{J.}~\bibnamefont{Liu}},
  \bibinfo{author}{\bibfnamefont{X.-P.} \bibnamefont{Wang}}, \bibnamefont{and}
  \bibinfo{author}{\bibfnamefont{W.}~\bibnamefont{Xue}} (\bibinfo{year}{2015}),
  \eprint{1512.07885}.

\bibitem[{\citenamefont{Cheung et~al.}(2015)\citenamefont{Cheung, Ko, Lee,
  Park, and Tseng}}]{Cheung:2015cug}
\bibinfo{author}{\bibfnamefont{K.}~\bibnamefont{Cheung}},
  \bibinfo{author}{\bibfnamefont{P.}~\bibnamefont{Ko}},
  \bibinfo{author}{\bibfnamefont{J.~S.} \bibnamefont{Lee}},
  \bibinfo{author}{\bibfnamefont{J.}~\bibnamefont{Park}}, \bibnamefont{and}
  \bibinfo{author}{\bibfnamefont{P.-Y.} \bibnamefont{Tseng}}
  (\bibinfo{year}{2015}), \eprint{1512.07853}.

\bibitem[{\citenamefont{Das and Rai}(2015)}]{Das:2015enc}
\bibinfo{author}{\bibfnamefont{K.}~\bibnamefont{Das}} \bibnamefont{and}
  \bibinfo{author}{\bibfnamefont{S.~K.} \bibnamefont{Rai}}
  (\bibinfo{year}{2015}), \eprint{1512.07789}.

\bibitem[{\citenamefont{Craig et~al.}(2015)\citenamefont{Craig, Draper, Kilic,
  and Thomas}}]{Craig:2015lra}
\bibinfo{author}{\bibfnamefont{N.}~\bibnamefont{Craig}},
  \bibinfo{author}{\bibfnamefont{P.}~\bibnamefont{Draper}},
  \bibinfo{author}{\bibfnamefont{C.}~\bibnamefont{Kilic}}, \bibnamefont{and}
  \bibinfo{author}{\bibfnamefont{S.}~\bibnamefont{Thomas}}
  (\bibinfo{year}{2015}), \eprint{1512.07733}.

\bibitem[{\citenamefont{Davoudiasl and Zhang}(2015)}]{Davoudiasl:2015cuo}
\bibinfo{author}{\bibfnamefont{H.}~\bibnamefont{Davoudiasl}} \bibnamefont{and}
  \bibinfo{author}{\bibfnamefont{C.}~\bibnamefont{Zhang}}
  (\bibinfo{year}{2015}), \eprint{1512.07672}.

\bibitem[{\citenamefont{Allanach et~al.}(2015)\citenamefont{Allanach, Dev,
  Renner, and Sakurai}}]{Allanach:2015ixl}
\bibinfo{author}{\bibfnamefont{B.~C.} \bibnamefont{Allanach}},
  \bibinfo{author}{\bibfnamefont{P.~S.~B.} \bibnamefont{Dev}},
  \bibinfo{author}{\bibfnamefont{S.~A.} \bibnamefont{Renner}},
  \bibnamefont{and} \bibinfo{author}{\bibfnamefont{K.}~\bibnamefont{Sakurai}}
  (\bibinfo{year}{2015}), \eprint{1512.07645}.

\bibitem[{\citenamefont{Salvio and Mazumdar}(2016)}]{Salvio:2015jgu}
\bibinfo{author}{\bibfnamefont{A.}~\bibnamefont{Salvio}} \bibnamefont{and}
  \bibinfo{author}{\bibfnamefont{A.}~\bibnamefont{Mazumdar}},
  \bibinfo{journal}{Phys. Lett.} \textbf{\bibinfo{volume}{B755}},
  \bibinfo{pages}{469} (\bibinfo{year}{2016}), \eprint{1512.08184}.

\bibitem[{\citenamefont{Aad et~al.}(2015)}]{Aad:2015owa}
\bibinfo{author}{\bibfnamefont{G.}~\bibnamefont{Aad}} \bibnamefont{et~al.}
  (\bibinfo{collaboration}{ATLAS}), \bibinfo{journal}{JHEP}
  \textbf{\bibinfo{volume}{12}}, \bibinfo{pages}{055} (\bibinfo{year}{2015}),
  \eprint{1506.00962}.

\bibitem[{\citenamefont{Khachatryan
  et~al.}(2014{\natexlab{a}})}]{Khachatryan:2014hpa}
\bibinfo{author}{\bibfnamefont{V.}~\bibnamefont{Khachatryan}}
  \bibnamefont{et~al.} (\bibinfo{collaboration}{CMS}), \bibinfo{journal}{JHEP}
  \textbf{\bibinfo{volume}{08}}, \bibinfo{pages}{173}
  (\bibinfo{year}{2014}{\natexlab{a}}), \eprint{1405.1994}.

\bibitem[{\citenamefont{Khachatryan
  et~al.}(2014{\natexlab{b}})}]{Khachatryan:2014gha}
\bibinfo{author}{\bibfnamefont{V.}~\bibnamefont{Khachatryan}}
  \bibnamefont{et~al.} (\bibinfo{collaboration}{CMS}), \bibinfo{journal}{JHEP}
  \textbf{\bibinfo{volume}{08}}, \bibinfo{pages}{174}
  (\bibinfo{year}{2014}{\natexlab{b}}), \eprint{1405.3447}.

\bibitem[{\citenamefont{Frampton}(1992)}]{Frampton:1992wt}
\bibinfo{author}{\bibfnamefont{P.~H.} \bibnamefont{Frampton}},
  \bibinfo{journal}{Phys. Rev. Lett.} \textbf{\bibinfo{volume}{69}},
  \bibinfo{pages}{2889} (\bibinfo{year}{1992}).

\bibitem[{\citenamefont{Pisano and Pleitez}(1992)}]{Pisano:1991ee}
\bibinfo{author}{\bibfnamefont{F.}~\bibnamefont{Pisano}} \bibnamefont{and}
  \bibinfo{author}{\bibfnamefont{V.}~\bibnamefont{Pleitez}},
  \bibinfo{journal}{Phys. Rev.} \textbf{\bibinfo{volume}{D46}},
  \bibinfo{pages}{410} (\bibinfo{year}{1992}), \eprint{hep-ph/9206242}.

\bibitem[{\citenamefont{Dias and Pleitez}(2004)}]{Dias:2003zt}
\bibinfo{author}{\bibfnamefont{A.~G.} \bibnamefont{Dias}} \bibnamefont{and}
  \bibinfo{author}{\bibfnamefont{V.}~\bibnamefont{Pleitez}},
  \bibinfo{journal}{Phys. Rev.} \textbf{\bibinfo{volume}{D69}},
  \bibinfo{pages}{077702} (\bibinfo{year}{2004}), \eprint{hep-ph/0308037}.

\bibitem[{\citenamefont{Dias et~al.}(2003)\citenamefont{Dias, de~S.~Pires, and
  da~Silva}}]{Dias:2003iq}
\bibinfo{author}{\bibfnamefont{A.~G.} \bibnamefont{Dias}},
  \bibinfo{author}{\bibfnamefont{C.~A.} \bibnamefont{de~S.~Pires}},
  \bibnamefont{and} \bibinfo{author}{\bibfnamefont{P.~S.~R.}
  \bibnamefont{da~Silva}}, \bibinfo{journal}{Phys. Rev.}
  \textbf{\bibinfo{volume}{D68}}, \bibinfo{pages}{115009}
  (\bibinfo{year}{2003}), \eprint{hep-ph/0309058}.

\bibitem[{\citenamefont{Diaz et~al.}(2005)\citenamefont{Diaz, Martinez, and
  Ochoa}}]{Diaz:2004fs}
\bibinfo{author}{\bibfnamefont{R.~A.} \bibnamefont{Diaz}},
  \bibinfo{author}{\bibfnamefont{R.}~\bibnamefont{Martinez}}, \bibnamefont{and}
  \bibinfo{author}{\bibfnamefont{F.}~\bibnamefont{Ochoa}},
  \bibinfo{journal}{Phys. Rev.} \textbf{\bibinfo{volume}{D72}},
  \bibinfo{pages}{035018} (\bibinfo{year}{2005}), \eprint{hep-ph/0411263}.

\bibitem[{\citenamefont{Dias et~al.}(2005{\natexlab{a}})\citenamefont{Dias,
  Martinez, and Pleitez}}]{Dias:2004dc}
\bibinfo{author}{\bibfnamefont{A.~G.} \bibnamefont{Dias}},
  \bibinfo{author}{\bibfnamefont{R.}~\bibnamefont{Martinez}}, \bibnamefont{and}
  \bibinfo{author}{\bibfnamefont{V.}~\bibnamefont{Pleitez}},
  \bibinfo{journal}{Eur. Phys. J.} \textbf{\bibinfo{volume}{C39}},
  \bibinfo{pages}{101} (\bibinfo{year}{2005}{\natexlab{a}}),
  \eprint{hep-ph/0407141}.

\bibitem[{\citenamefont{Dias}(2005)}]{Dias:2004wk}
\bibinfo{author}{\bibfnamefont{A.~G.} \bibnamefont{Dias}},
  \bibinfo{journal}{Phys. Rev.} \textbf{\bibinfo{volume}{D71}},
  \bibinfo{pages}{015009} (\bibinfo{year}{2005}), \eprint{hep-ph/0412163}.

\bibitem[{\citenamefont{Dias et~al.}(2005{\natexlab{b}})\citenamefont{Dias,
  Doff, de~S.~Pires, and Rodrigues~da Silva}}]{Dias:2005jm}
\bibinfo{author}{\bibfnamefont{A.~G.} \bibnamefont{Dias}},
  \bibinfo{author}{\bibfnamefont{A.}~\bibnamefont{Doff}},
  \bibinfo{author}{\bibfnamefont{C.~A.} \bibnamefont{de~S.~Pires}},
  \bibnamefont{and} \bibinfo{author}{\bibfnamefont{P.~S.}
  \bibnamefont{Rodrigues~da Silva}}, \bibinfo{journal}{Phys. Rev.}
  \textbf{\bibinfo{volume}{D72}}, \bibinfo{pages}{035006}
  (\bibinfo{year}{2005}{\natexlab{b}}), \eprint{hep-ph/0503014}.

\bibitem[{\citenamefont{Dias et~al.}(2005{\natexlab{c}})\citenamefont{Dias,
  de~S.~Pires, and Rodrigues~da Silva}}]{Dias:2005yh}
\bibinfo{author}{\bibfnamefont{A.~G.} \bibnamefont{Dias}},
  \bibinfo{author}{\bibfnamefont{C.~A.} \bibnamefont{de~S.~Pires}},
  \bibnamefont{and} \bibinfo{author}{\bibfnamefont{P.~S.}
  \bibnamefont{Rodrigues~da Silva}}, \bibinfo{journal}{Phys. Lett.}
  \textbf{\bibinfo{volume}{B628}}, \bibinfo{pages}{85}
  (\bibinfo{year}{2005}{\natexlab{c}}), \eprint{hep-ph/0508186}.

\bibitem[{\citenamefont{Dias et~al.}(2011)\citenamefont{Dias, de~S.~Pires, and
  da~Silva}}]{Dias:2011sq}
\bibinfo{author}{\bibfnamefont{A.~G.} \bibnamefont{Dias}},
  \bibinfo{author}{\bibfnamefont{C.~A.} \bibnamefont{de~S.~Pires}},
  \bibnamefont{and} \bibinfo{author}{\bibfnamefont{P.~S.~R.}
  \bibnamefont{da~Silva}}, \bibinfo{journal}{Phys. Rev.}
  \textbf{\bibinfo{volume}{D84}}, \bibinfo{pages}{053011}
  (\bibinfo{year}{2011}), \eprint{1107.0739}.

\bibitem[{\citenamefont{Ochoa and Martinez}(2005)}]{Ochoa:2005ch}
\bibinfo{author}{\bibfnamefont{F.}~\bibnamefont{Ochoa}} \bibnamefont{and}
  \bibinfo{author}{\bibfnamefont{R.}~\bibnamefont{Martinez}},
  \bibinfo{journal}{Phys. Rev.} \textbf{\bibinfo{volume}{D72}},
  \bibinfo{pages}{035010} (\bibinfo{year}{2005}), \eprint{hep-ph/0505027}.

\bibitem[{\citenamefont{Peccei and Quinn}(1977{\natexlab{a}})}]{Peccei:1977hh}
\bibinfo{author}{\bibfnamefont{R.~D.} \bibnamefont{Peccei}} \bibnamefont{and}
  \bibinfo{author}{\bibfnamefont{H.~R.} \bibnamefont{Quinn}},
  \bibinfo{journal}{Phys. Rev. Lett.} \textbf{\bibinfo{volume}{38}},
  \bibinfo{pages}{1440} (\bibinfo{year}{1977}{\natexlab{a}}).

\bibitem[{\citenamefont{Peccei and Quinn}(1977{\natexlab{b}})}]{Peccei:1977ur}
\bibinfo{author}{\bibfnamefont{R.~D.} \bibnamefont{Peccei}} \bibnamefont{and}
  \bibinfo{author}{\bibfnamefont{H.~R.} \bibnamefont{Quinn}},
  \bibinfo{journal}{Phys. Rev.} \textbf{\bibinfo{volume}{D16}},
  \bibinfo{pages}{1791} (\bibinfo{year}{1977}{\natexlab{b}}).

\bibitem[{\citenamefont{Buras et~al.}(2013)\citenamefont{Buras, De~Fazio,
  Girrbach, and Carlucci}}]{Buras:2012dp}
\bibinfo{author}{\bibfnamefont{A.~J.} \bibnamefont{Buras}},
  \bibinfo{author}{\bibfnamefont{F.}~\bibnamefont{De~Fazio}},
  \bibinfo{author}{\bibfnamefont{J.}~\bibnamefont{Girrbach}}, \bibnamefont{and}
  \bibinfo{author}{\bibfnamefont{M.~V.} \bibnamefont{Carlucci}},
  \bibinfo{journal}{JHEP} \textbf{\bibinfo{volume}{02}}, \bibinfo{pages}{023}
  (\bibinfo{year}{2013}), \eprint{1211.1237}.

\bibitem[{\citenamefont{Cabarcas et~al.}(2008)\citenamefont{Cabarcas,
  Gomez~Dumm, and Martinez}}]{Cabarcas:2008ys}
\bibinfo{author}{\bibfnamefont{J.~M.} \bibnamefont{Cabarcas}},
  \bibinfo{author}{\bibfnamefont{D.}~\bibnamefont{Gomez~Dumm}},
  \bibnamefont{and} \bibinfo{author}{\bibfnamefont{R.}~\bibnamefont{Martinez}},
  \bibinfo{journal}{Eur. Phys. J.} \textbf{\bibinfo{volume}{C58}},
  \bibinfo{pages}{569} (\bibinfo{year}{2008}), \eprint{0809.0821}.

\bibitem[{\citenamefont{Ky and Van}(2005)}]{Ky:2005yq}
\bibinfo{author}{\bibfnamefont{N.~A.} \bibnamefont{Ky}} \bibnamefont{and}
  \bibinfo{author}{\bibfnamefont{N.~T.~H.} \bibnamefont{Van}},
  \bibinfo{journal}{Phys. Rev.} \textbf{\bibinfo{volume}{D72}},
  \bibinfo{pages}{115017} (\bibinfo{year}{2005}), \eprint{hep-ph/0512096}.

\bibitem[{\citenamefont{Djouadi}(2008)}]{Djouadi:2005gi}
\bibinfo{author}{\bibfnamefont{A.}~\bibnamefont{Djouadi}},
  \bibinfo{journal}{Phys. Rept.} \textbf{\bibinfo{volume}{457}},
  \bibinfo{pages}{1} (\bibinfo{year}{2008}), \eprint{hep-ph/0503172}.

\bibitem[{\citenamefont{Shifman et~al.}(1979)\citenamefont{Shifman, Vainshtein,
  Voloshin, and Zakharov}}]{Shifman:1979eb}
\bibinfo{author}{\bibfnamefont{M.~A.} \bibnamefont{Shifman}},
  \bibinfo{author}{\bibfnamefont{A.~I.} \bibnamefont{Vainshtein}},
  \bibinfo{author}{\bibfnamefont{M.~B.} \bibnamefont{Voloshin}},
  \bibnamefont{and} \bibinfo{author}{\bibfnamefont{V.~I.}
  \bibnamefont{Zakharov}}, \bibinfo{journal}{Sov. J. Nucl. Phys.}
  \textbf{\bibinfo{volume}{30}}, \bibinfo{pages}{711} (\bibinfo{year}{1979}),
  \bibinfo{note}{[Yad. Fiz.30,1368(1979)]}.

\bibitem[{\citenamefont{Cao et~al.}(2010)\citenamefont{Cao, Jackson, Keung,
  Low, and Shu}}]{Cao:2009ah}
\bibinfo{author}{\bibfnamefont{Q.-H.} \bibnamefont{Cao}},
  \bibinfo{author}{\bibfnamefont{C.~B.} \bibnamefont{Jackson}},
  \bibinfo{author}{\bibfnamefont{W.-Y.} \bibnamefont{Keung}},
  \bibinfo{author}{\bibfnamefont{I.}~\bibnamefont{Low}}, \bibnamefont{and}
  \bibinfo{author}{\bibfnamefont{J.}~\bibnamefont{Shu}},
  \bibinfo{journal}{Phys. Rev.} \textbf{\bibinfo{volume}{D81}},
  \bibinfo{pages}{015010} (\bibinfo{year}{2010}), \eprint{0911.3398}.

\bibitem[{\citenamefont{Low et~al.}(2010)\citenamefont{Low, Rattazzi, and
  Vichi}}]{Low:2009di}
\bibinfo{author}{\bibfnamefont{I.}~\bibnamefont{Low}},
  \bibinfo{author}{\bibfnamefont{R.}~\bibnamefont{Rattazzi}}, \bibnamefont{and}
  \bibinfo{author}{\bibfnamefont{A.}~\bibnamefont{Vichi}},
  \bibinfo{journal}{JHEP} \textbf{\bibinfo{volume}{04}}, \bibinfo{pages}{126}
  (\bibinfo{year}{2010}), \eprint{0907.5413}.

\bibitem[{\citenamefont{Aad et~al.}(2014)}]{Aad:2014fha}
\bibinfo{author}{\bibfnamefont{G.}~\bibnamefont{Aad}} \bibnamefont{et~al.}
  (\bibinfo{collaboration}{ATLAS}), \bibinfo{journal}{Phys. Lett.}
  \textbf{\bibinfo{volume}{B738}}, \bibinfo{pages}{428} (\bibinfo{year}{2014}),
  \eprint{1407.8150}.

\bibitem[{\citenamefont{Cao et~al.}(2015{\natexlab{d}})\citenamefont{Cao, Yan,
  and Zhang}}]{Cao:2015lia}
\bibinfo{author}{\bibfnamefont{Q.-H.} \bibnamefont{Cao}},
  \bibinfo{author}{\bibfnamefont{B.}~\bibnamefont{Yan}}, \bibnamefont{and}
  \bibinfo{author}{\bibfnamefont{D.-M.} \bibnamefont{Zhang}},
  \bibinfo{journal}{Phys. Rev.} \textbf{\bibinfo{volume}{D92}},
  \bibinfo{pages}{095025} (\bibinfo{year}{2015}{\natexlab{d}}),
  \eprint{1507.00268}.

\end{thebibliography}

\end{document}